\documentclass[prd,nofootinbib,preprint,superscriptaddress]{revtex4-2}

\usepackage{amsmath, amssymb, amsthm, graphicx, epsfig, fancyhdr,epsfig, slashed, mathrsfs}

\usepackage{tikzsymbols}
\usepackage{natbib}
\usepackage{float}
\usepackage{graphicx}
\usepackage{amsmath}
\usepackage{amsfonts}
\usepackage{amssymb}
\usepackage[bookmarks=true,bookmarksopen=true,
                    bookmarksnumbered=true,
                    colorlinks=true,linkcolor=black,
                    citecolor=red]{hyperref}
\usepackage{subfigure}
\usepackage{slashed}
 \usepackage{setspace} 
\usepackage{caption}
\definecolor{lime}{HTML}{A6CE39}
\DeclareRobustCommand{\orcidicon}{
	\begin{tikzpicture}
		\draw[lime, fill=lime] (0,0) 
		circle [radius=0.2] 
		node[white] {{\fontfamily{qag}\selectfont \tiny ID}};
		\draw[white, fill=white] (-0.0625,0.095) 
		circle [radius=0.007];
	\end{tikzpicture}
	\hspace{-2mm}
}

\foreach \x in {A, ..., Z}{\expandafter\xdef\csname orcid\x\endcsname{\noexpand\href{https://orcid.org/\csname orcidauthor\x\endcsname}
		{\noexpand\orcidicon}}
}

\newcommand{\be}{\begin{equation}}
	\newcommand{\ee}{\end{equation}}
\newcommand{\bea}{\begin{eqnarray}}
	\newcommand{\eea}{\end{eqnarray}}

\newcommand{\beq}{\begin{equation}}
	\newcommand{\eeq}{\end{equation}}

\def\nn{\nonumber}

\def\mpl{M_{\rm Pl}}

\usepackage[utf8]{inputenc}

\DeclareUnicodeCharacter{2212}{-}

\begin{document}
\hspace*{\fill} HRI-RECAPP-2022-011

\title{The stability analysis of the extended singlet scalar model with two high scale minima}

\author{\small Najimuddin Khan \orcidC{}}
\email{najimuddinkhan@hri.res.in }
\affiliation{\small Harish-Chandra Research Institute, A CI of Homi Bhabha National Institute, Chhatnag Road, Jhunsi, Prayagraj 211019, India\vspace{1.80cm}}

\begin{abstract}
We study the simplest viable dark matter (DM) model with a real singlet scalar, vector-like singlet, and doublet fermions.
The Yukawa couplings associated with the fermion sector are important in obtaining the current DM relic density through Freeze-out and Freeze-in mechanisms.
In addition to the standard model-like high scale minima along the Higgs field, we find other high scale minima along the singlet scalar field direction.
The other minima appears due to the renormalisation group evolutions of the couplings involving the gauge sector, Higgs portal, and the new Yukawa sector.
A detailed study of the parameter space identifying the region of electroweak vacuum stability and metastability along different directions of the scalar field is carried out using different phase diagrams.
\end{abstract}
	\maketitle

\section{Introduction}

The standard model (SM) of particle physics is indeed one of the most successful theories of the last century; however, it still has a few unsolved empirical observations. The strong CP problem, neutrino mass and mixing, matter-antimatter asymmetry, nature of dark matter (DM) dark energy, and inflation-related issues have been considered shortfalls of the SM. The last few decades have seen an upheaval in astrophysics and cosmology. The Universe is filled with matter, dark matter, and a large amount of dark energy.
From the standard model model point of view, there is no sufficient candidate left to propose as a dark matter candidate. Therefore, one must go beyond the standard model to explain the current dark matter density.
The recent LHC Higgs signal strength data~\cite{Sirunyan:2017khh, Sirunyan:2018koj} also allows us to include additional fields of the new physics beyond the SM.

The standard model theory is also viewed as unnatural if a severe fine-tuning between the quadratic radiative corrections~\cite{Susskind:1978ms,tHooft:1980xss, Giudice:2013yca}. A bare mass is required to return the Higg mass to the observed scale.
It is also known that dimensional regularization can remove the quadratic divergences. However, other logarithmic and finite contributions can cause a similar naturalness problem~\cite{Vissani:1997ys, Giudice:2013yca}.
It is also worth noting that the $125.5 \pm 0.3$ GeV Higgs mass observed at the LHC is quite intriguing from a vacuum stability perspective~\cite{Giudice:2013yca, Vissani:1997ys, Buttazzo:2013uya, Degrassi:2012ry}.
For Higgs field values $h>>v_{SM}=246.221$ GeV, the Higgs effective potential~\cite{Giudice:2013yca,Vissani:1997ys,Buttazzo:2013uya,Degrassi:2012ry,Alekhin:2012py} turns down due to large contributions from top quark in loops.
The observed values of the standard model parameters, especially the top mass and strong coupling constant, suggest an extra deeper minima situated near the Planck scale. The presence of this high scale deeper minima is a threatening situation for the stability of the present electroweak vacuum, as the Universe may tunnel into that true (deeper) vacuum.
We know the tunneling probability, calculated using the state-of-the-art next-to-next-to-leading- order corrections. We find that the present EW
vacuum is metastable at $3\sigma$, i.e., the decay time is greater than the age of the Universe, which demands the new physics at a low and/or high scale. The observed dark matter and neutrino mass could be interesting to explore in order to solve these issues~\cite{Khan:2014kba,Khan:2015ipa,Khan:2016sxm,Garg:2017iva}.

One can get the exact relic density of the dark matter using various theories~\cite{Kolb:1990vq, Hall:2009bx}.
Many of the dark matter genesis theories are based upon the thermal Freeze-out mechanism. In this Freeze-out mechanism, the weakly interacting massive particles (WIMPs) are in thermal equilibrium in the early universe, i.e., the dark matter has a large initial density equal to the other particle densities.
When the primordial temperature drops below the mass of the DM, $i.e.$, $T<M_{DM}$, it dilutes away until the annihilation and/or co-annihilation into the lighter particles becomes slower than the expansion rate of the universe, the co-moving dark matter number density becomes constant.
We also have seen that most of the single component WIMP dark matter models are tightly constrained by the present direct detection constraints~\cite{Aprile:2018dbl}.
The low mass region may explain the right relic density $\Omega h^2=0.1198\pm 0.0026$~\cite{Aghanim:2018eyx}, however, the direct detection data exclude most of them~\cite{Burgess:2000yq,Deshpande:1977rw}.

In this study, the standard model is extended with a real singlet scalar, and singlet and doublet fermions~\cite{Das:2020hpd, Das:2021zea} is revisited in the context of the low and high dark matter mass region. 
We have already shown a considerable increment in the dark matter (as WIMP and FIMP) parameter spaces in the presence of the new Yukawa couplings~\cite{Das:2020hpd, Das:2021zea}. We reduced the Higgs portal coupling $\kappa$ to avoid the direct detection limits.
The additional $t,u$-channels through the new fermions can increase the dark matter annihilation cross-section to produce the exact relic density. The relic density can be achieved for a large region of the parameter space by adjusting the Higgs portal and new Yukawa couplings. These couplings also have an impact on the neutrino low energy variables.
Now the questions are (a) how large/small of these couplings can be taken at the electroweak scale? (b) what happens at a very high scale? (c) Can these new Yukawa couplings aggravate large-scale minima along Higgs' direction? (d) Can we have more minima in different scalar field directions?

In this minimal setup of our model work, we discuss the above questions. 
In the present work, we identify the parameter space relevant to low and heavy dark matter mass regions, including a detailed stability analysis.
We know that depending on the coupling strength(s), the additional new scalar pulls the electroweak vacuum towards stability whereas the fermion push it towards instability~\cite{Khan:2016sxm,Rodejohann:2012px,Chakrabortty:2012np,Khan:2012zw,Datta:2013mta,Chakrabortty:2013zja,Kobakhidze:2013pya,Garg:2017iva,Goswami:2018jar,Khan:2015ipa,Khan:2014kba}.
Thus, in our case, the Higgs portal, new Yukawa couplings, including the masses of the additional scalar and fermions, will also get
bounded by the constraints from the stability/metastability
of the electroweak vacuum.
So far, from a stability viewpoint, we have seen in the literature that there could be a minima only along the Higgs field direction due to the renormalisation group evolutions of the couplings involving the gauge sector, Yukawa sector, and Higgs quartic couplings. We find a similar situation to the additional scalar field direction due to the presence of new Yukawa, Higgs portal, and gauge couplings (mainly $g_1$ and $g_2$). We calculate the tunneling probability for both the high scale minima and discuss the detailed analysis in this work.
Depending on the depth of these minima, the tunneling probability will change. The current electroweak vacuum only tends to tunnel to the deeper minima (either in the Higgs or scalar directions). If the depth of these new minima is the same, then electroweak vacuum tunnel to the nearest one only. We can also find a situation where the depth of these two high scale minima and the length (bounce scale) from the electroweak minima is the same. However, the dark matter, neutrino low energy variable, and other constraints will exclude those parameter spaces.
To the best of our knowledge, the stability bounds of this model (or in a similar model like this), which has two or more minima at a high scale, were not discussed in the literature.

The rest of the work is organized as follows. We have given the complete model description in section~\ref{sec2}.
The possible constraints in section~\ref{sec:lfv}. The dark matter analysis through Freeze-out and Freeze-in mechanism is carried out in detail in section~\ref{dm1}. The detailed stability analysis is carried out in section~\ref{sec:stab}. We briefly talk about inflation and reheating in section~\ref{sec:inflation}. 
Finally, we lay down our conclusions in section~\ref{conc}.
 
\section{Model}\label{sec2}
The model addressed here, contains (i) a real singlet scalar $S$, (ii) a vector-like charged fermion singlet $E_S^-$ and (iii) a vector-like fermion doublet, $F_D=(X_1^0~~E_D^-)^T$.
It is to be noted that these additional fermions are vector-like, and hence, they do not introduce any new anomalies ~\cite{pal2014introductory, Pisano:1993es}.
All the newly added particles are considered odd under discrete $Z_2$ symmetry ($\Psi\rightarrow-\Psi$), such that these fields do not mix with the standard model fields. Hence, the lightest and neutral particle is stable and can serve as a viable dark matter candidate. The Lagrangian of the model read as~\cite{Das:2020hpd,Das:2021zea},
\begin{equation}
	\mathcal{L}=\mathcal{L_{\rm SM}}+\mathcal{L_S}+\mathcal{L_F}+\mathcal{L}_{int},
\end{equation}
where,
\begin{eqnarray}
	\mathcal{L_S}&=&\frac{1}{2}|\partial_{\mu}S|^2-\frac{1}{2}kS^2\phi^2-\frac{1}{4}m_{S}^2S^2-\frac{\lambda_S}{4!}S^4\label{eq:scalar},\\
	\nn \mathcal{L_F}&=&\overline{F}_D\gamma^{\mu}D_{\mu}F_D+\overline{E}_S\gamma^{\mu}D_{\mu} E_S-M_{ND}\overline{F}_DF_D - M_{NS}\overline{E}_SE_S,\\
	\mathcal{L}_{int}&=&-Y_{N}\overline{F}_D\phi E_S - Y_{fi} \overline{\psi}_{i,L} F_D S  - \, Y_{fi}^{\prime }\, \overline{l}_{i,R} E_S S + h.c.~\label{lint}
\end{eqnarray}
Here, $D_{\mu}$ stands for the corresponding covariant derivative of the doublet and singlet fermions. The left-handed lepton doublet is denoted by ${\psi}_{i,L}=(\nu_i, l_i)_L^T$, whereas $l_{i,R}$ stand for the right-handed singlet charged leptons, with indices $i=e,\mu,\tau$ three generation of leptons. $L$ and $R$ indicate left and right chirality of the fermions.
The standard model Higgs potential can be written as,  $V^{SM}(\phi)=-m^2\phi^2+\lambda\phi^4$, where,  
$\phi=( G^+, \frac{H+v+iG}{\sqrt{2}})^T$  is the standard model Higgs doublet. The Goldstone bosons denote as $G$ and $v=246.221$ GeV is the vacuum expectation value of the Higgs $H$ scalar field. The mass matrix for these charged fermion fields can be written as,
\begin{eqnarray}
	\mathcal{M}=\begin{pmatrix}
		M_{ND}&M_X\\M_X^{\dagger}&M_{NS}\\
	\end{pmatrix},
	\label{eq:mass1}
\end{eqnarray}
with, $M_X=\frac{Y_{N}v}{\sqrt{2}}$. The charged component of the fermion doublet $E_D^\pm$ and the singlet charged fermion ($E_S^\pm$) mix at tree level. 
The mass eigenvalues and eigenstates are obtained by diagonalizing the mass matrix with
a rotation of the ($E_D^\pm$  $E_S^\pm$) basis,
\begin{eqnarray}
	\begin{pmatrix}
		E_1^\pm\\E_2^\pm\\
	\end{pmatrix}=\begin{pmatrix}
		\cos\beta&\sin\beta\\-\sin\beta&\cos\beta\\
	\end{pmatrix}\begin{pmatrix}
		E_D^\pm\\E_S^\pm\\
	\end{pmatrix}, {~\rm with~} \tan 2 \beta = \frac{2 M_X}{M_{NS}-M_{ND}}.
\end{eqnarray}
Diagonalization of eqn.~\ref{eq:mass1} can give the following eigenvalues for the charged leptons ($M_{NS}-M_{ND} \gg M_X$) as,
\begin{eqnarray}
	M_{E_1^\pm} = M_{ND} - \frac{2 (M_X)^2}{M_{NS}-M_{ND}},\,
	M_{E_2^\pm} = M_{NS} + \frac{2 (M_X)^2}{M_{NS}-M_{ND}}.\nn
\end{eqnarray}
The masses of the neutral fermion and scalar fields can be written as,
\begin{eqnarray}
	M_{X_1^0}=M_{ND}, \, M_{S}^2=\frac{m_{S}^2+kv^2}{2}~ \ \text{and}~\ M_H^2= 2\lambda v^2. 
	\label{eq:mass}
\end{eqnarray} 
In this model, one can see that the neutral fermion can not be the dark matter candidate as $M_{E_1^\pm}<M_{X_1^0}<M_{E_2^\pm}$. Only the singlet scalar fields $S$, for $M_{S} < M_{E_1^\pm}$, can serve as a viable dark matter candidate. Here, one can obtain dark matter through both Freeze-out and Freeze-in mechanisms.
We will show a detailed discussion on the new region of the allowed parameter spaces and the effect of the additional $Z_2$-odd fermion in the dark matter section~\ref{dm1}.

The parameter space of this model is bounded by various constraints arising from theoretical considerations like absolute vacuum stability, metastability, and unitarity of the scattering matrix, observation phenomenons like dark matter relic density. 
The recent measurements of the Higgs invisible decay width and signal strength at the LHC put additional constraints.
The dark matter requirement saturates the dark matter relic density all alone restricts the allowed parameter space considerably. These constraints are already discussed in our previous paper~\cite{Das:2020hpd,Das:2021zea}. We discuss the most stringent absolute stability, unitarity, and lepton flavor violation ($\mu\rightarrow e\gamma$), incorporating the new anomalous magnetic moment result in the next section.

\subsection{Absolute stability and unitary constraints}
The potential `bounded from below', i.e., absolute stability of the electroweak vacuum, indicates that the scalar potential would not become minus infinity in any direction of the scalar field. In this model, for the very large value of the scalar fields $H, S>>v$, the scalar potential in eqn.~(\ref{eq:scalar}) can be written as,
\begin{eqnarray}
\nn V(H,~S) = \frac{1}{4}\left\lbrace \sqrt{\lambda} H^2 - \sqrt{\frac{\lambda_S}{6}} S^2 \right\rbrace^2 + \frac{1}{4}\left\lbrace\kappa +   \sqrt{\frac{2 \lambda \lambda_S}{3}}\right\rbrace H^2 S^2.
\label{scalpotstability}
\end{eqnarray}
We are working on the unitary gauge. The absolute stability conditions  for the scalar potential are then given by,
\be
\lambda(\Lambda) > 0, \quad \lambda_S(\Lambda) > 0 \quad {\rm and} \quad \kappa(\Lambda) + \sqrt{\frac{2 \lambda(\Lambda) \lambda_S(\Lambda)}{3}} > 0.
\ee
Here, this model's coupling constants are evaluated at a scale $\Lambda$ using RG equations~\cite{Khan:2012zw}. 
It is to be noted that these conditions can modify if the scalar quartic couplings $\lambda$ and/or $\lambda_S$ become negative at some energy scale. In this situation, we need to carefully handle the stability constraints as shown in Ref.~\cite{Khan:2012zw}. In the context of this paper, we present the detailed analysis in section~\ref{sec:stab} which is our main aim of this work.

Again, for the radiatively improved Lagrangian of our model to be perturbative, we should follow~\cite{Lee:1977eg, Cynolter:2004cq},
\be
\lambda (\Lambda)\, 
 <\, \frac{4\pi}{3} \, ; \,\,\, |\kappa(\Lambda)|\, < \,8\pi \,;\,\,\, |\lambda_{S}(\Lambda)|\, < \, 8\pi.
 \label{eq:pert}
 \ee

The couplings of the scalar potential ($\lambda,\kappa$, and $\lambda_S$) of this model are constrained by the unitarity of the scattering matrix (S-matrix). At very high field values, one can obtain the S-matrix using various gauge boson-gauge boson, scalar-scalar, and gauge-scalar boson scatterings. Using the equivalence theorem, we get the S-matrix. The unitarity demands that the eigenvalues of the S-matrix should be less than $8\pi$. The unitary bounds for this model~\cite{Cynolter:2004cq}, 
\bea
\lambda(\Lambda) \leq 8 \pi ~{\rm and}~ \Big| 12 {\lambda(\Lambda)}+{\lambda_S(\Lambda)} \pm \sqrt{16 \kappa(\Lambda)^2+({\lambda_S(\Lambda)}-12 {\lambda(\Lambda))^2}}\Big| \leq 32 \pi.
\label{eq:unitary}
\eea
\subsection{Lepton flavor violation ($\mu\rightarrow e\gamma$) and anomalous magnetic moment}\label{sec:lfv}
In this model, we see that large Yukawa couplings ($Y_{fi} > 4\pi$) are required to explain the present lepton anomalous magnetic moment data. For example, one can get $\delta a_\mu=25.1\times10^{-11}$~\cite{Abi:2021gix} for $Y_{f2} =29.32$, $M_{S}=1000$ GeV, $M_{E_1^\pm}=1500$ GeV and  $M_{E_2^\pm}=3000$ GeV. This choice of couplings and masses violate the lepton flavor violation (LFV)~\cite{Baldini:2018nnn} BR$(\mu \rightarrow e\gamma) $ $< 4.2 \times 10^{−13}$ at $90\%$ CL as well as perturbative limits. 
If we assume non-zero $Y_{f2}=Y_{f2}^{\prime}$, then for $Y_{f2}\sim 20$ to get $\delta a_\mu=25.1\times10^{-10}$. We also need large $Y_{f1}=Y_{f1}^{\prime}$ to obtain electron anomalous magnetic moment.
It is to be noted that one can use the second generation Yukawa couplings at $Y_{f2} $ and $Y_{f2}^{\prime}< 10^{-3}$ and other couplings are $\mathcal{O}(1)$ to evade the LFV and anomalous magnetic moment constraints~\cite{Das:2020hpd, Das:2021zea}.
On the contrary, these large couplings are not good for this model's scalar potential as they will render a large negative potential (become unbounded from below) towards the singlet and Higgs scalar fields due to radiative corrections.  
We discuss these limits more carefully in the stability analysis and dark matter sections.

\section{Dark matter}
\label{dm1}
As pointed out in the previous analysis~\cite{Das:2020hpd, Das:2021zea}, the lightest $Z_2$-odd singlet scalar $S$ is the viable dark matter (DM) candidate. Here, the dark matter can give the proper relic density using the Freeze-out and Freeze-in mechanisms. The parameter spaces change with the choice of these processes.
If the dark matter is in thermal equilibrium in the early Universe, it can annihilate into the  standard model particles when $T>M_{DM}$. $T$ stands for the universe temperature. 
The dark matter Freezes out near $T<M_{DM}$ and gives dark matter density depending on the parameter spaces.
However, if the dark matter candidate was not in thermal equilibrium in the early Universe, it could produce from some other heavy particles (known as mother particles for dark matter). It then contributes to the current density via the Freeze-in mechanism.
Dark matter produced from decay and/or annihilation of various other particles is in thermal equilibrium in the early Universe. The condition can be written as follows, 
\beq
{\Gamma \over H(T) } \geq 1,
\eeq
where, $\Gamma$ is the relevant decay width and  $H(T)=\left( g^* \, \frac{\pi^2}{90} \, \frac{T^4}{\mpl^2} \right)^{1/2}$ is the Hubble parameter~\cite{Plehn:2017fdg,Hall:2009bx,Biswas:2016bfo} where, $\mpl\approx2.4\times 10^{18}$ GeV is the reduced Planck mass.
For the production of particles that occur mainly from the annihilation of other particles in the thermal bath, then  $\Gamma=n_{eq} <\sigma v>$~\cite{Plehn:2017fdg,Hall:2009bx,Biswas:2016bfo}. Here equilibrium number density is given by~\cite{Plehn:2017fdg},
\beq
\begin{aligned}
	&n_{eq} &=
	&\left\{ \begin{array}{l} \vspace{0.3cm} g^* \left( \frac{m T}{2 \pi}\right)^{3/2} \, e^{-m/T},  ~~~~~~~{\rm for ~non\text{-}relativistic ~states}~~T<<M\\
		
		\frac{\zeta_3}{\pi^2} g^* T^3,  ~~~~~~~~~~~~~~~~~~~~{\rm for~relativistic~boson ~states}~~T>>M\\
		
		\vspace{0.5cm}
		\frac{ 3}{4}\, \frac{\zeta_3}{\pi^2} g^* T^3 ,  ~~~~~~~~~~~~~~~~~~{\rm for~relativistic~fermion ~states}~~T>>M
		\vspace{-0.2cm}\end{array}
	\right.
	\label{eq:n}
\end{aligned}
\eeq
\begin{figure}[h!]
	\begin{center}
		\subfigure[]{
			\includegraphics[scale=0.6]{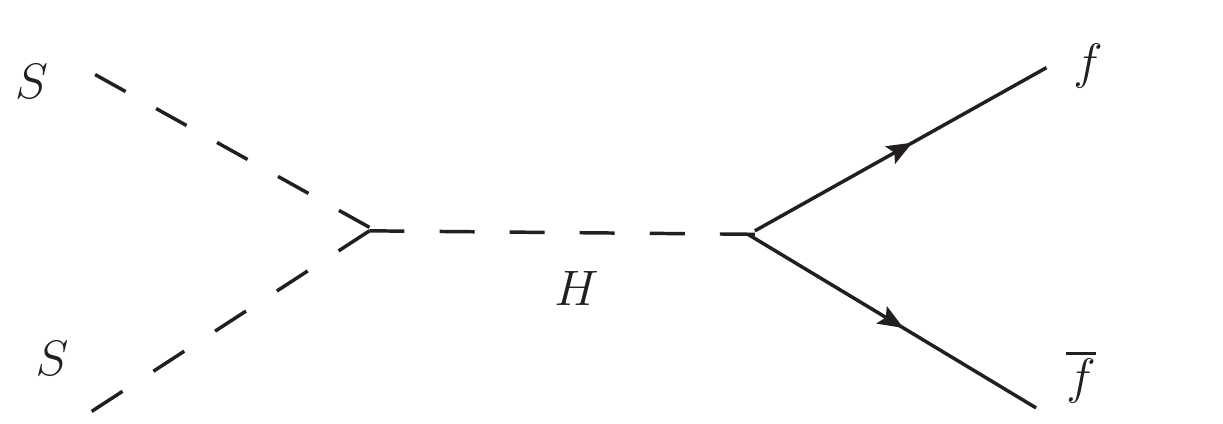}}
		\hskip 1pt
		\subfigure[]{
			\includegraphics[scale=0.6]{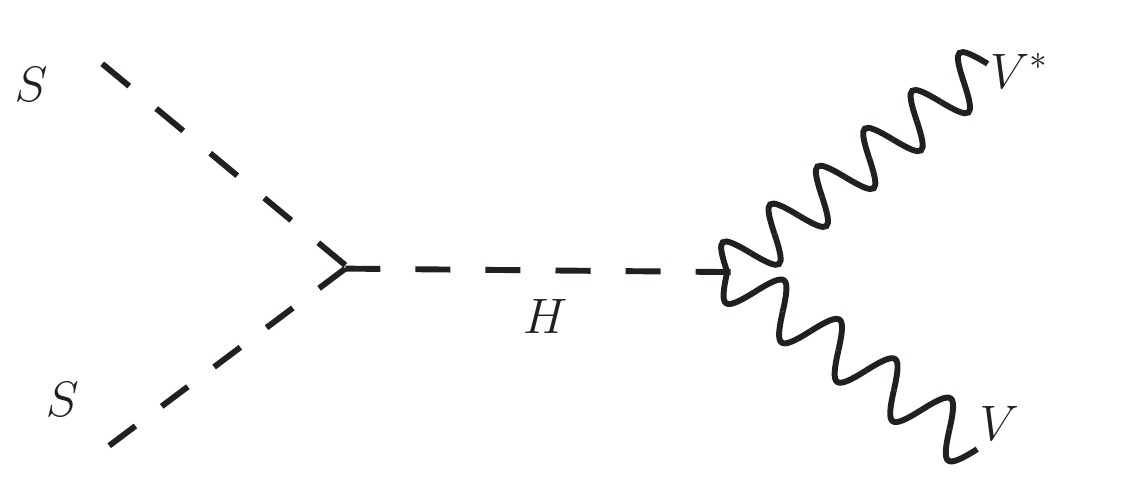}}
		\subfigure[]{
			\includegraphics[scale=0.6]{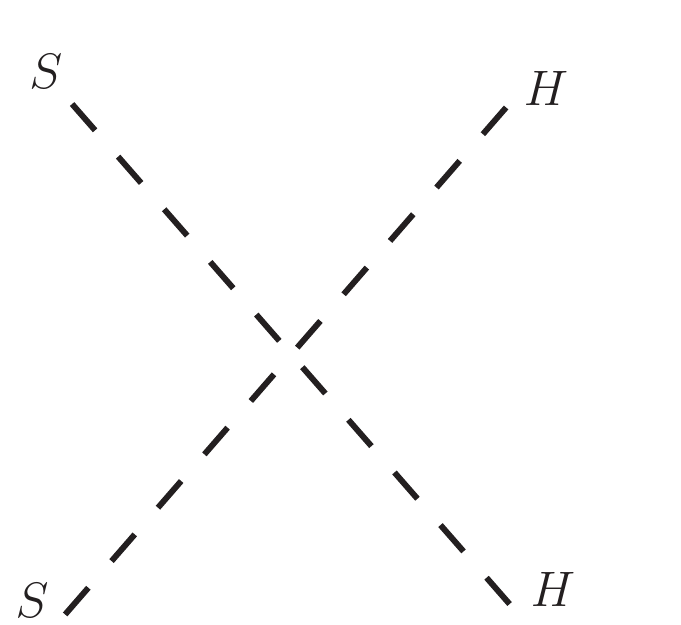}}
		\hskip 1pt
		\subfigure[]{
			\includegraphics[scale=0.6]{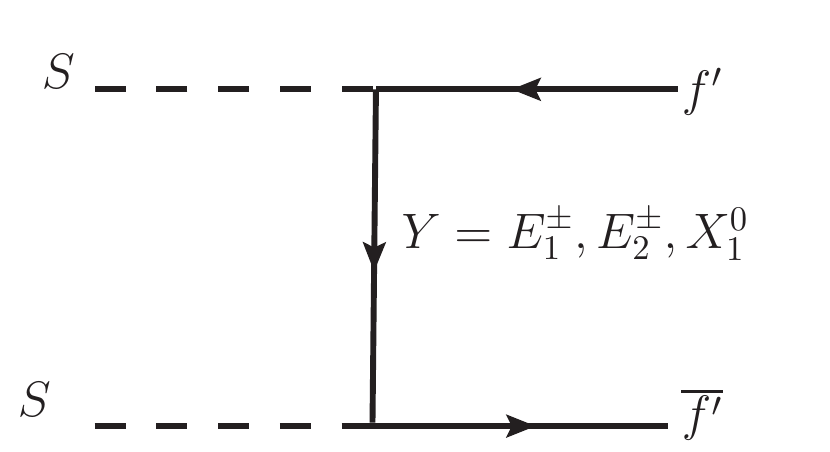}}
		\caption{ \it The dark matter annihilation diagrams give the relic density. $V$ stands for gauge bosons $W,Z$; $f'$ represents the standard model leptons and $f$ are standard model leptons and quarks.}
		\label{fig:DarkAn}
	\end{center}
\end{figure}

\begin{figure}[h!]
	\begin{center}
		\subfigure[]{
			\includegraphics[scale=0.6]{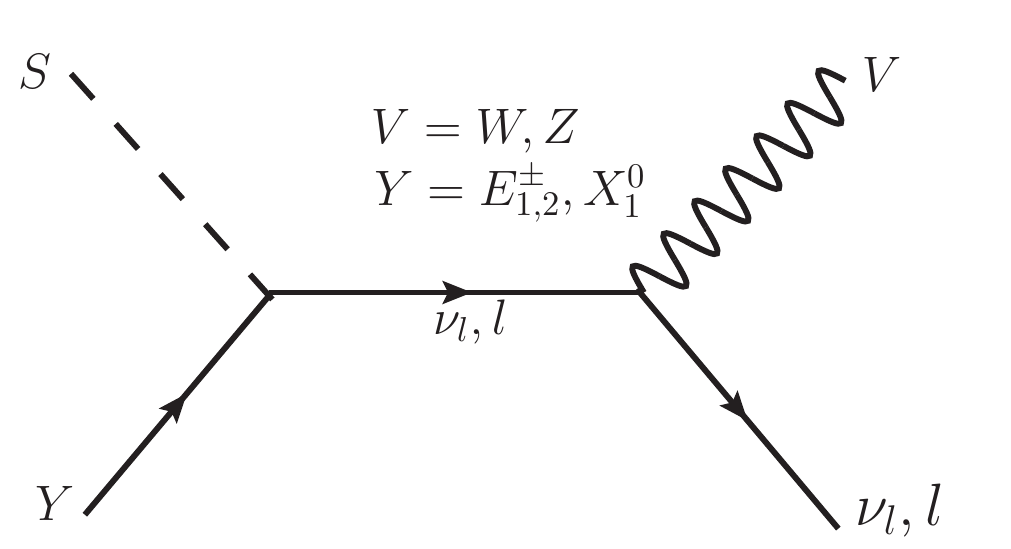}}
		\hskip 1pt
		\subfigure[]{
			\includegraphics[scale=0.6]{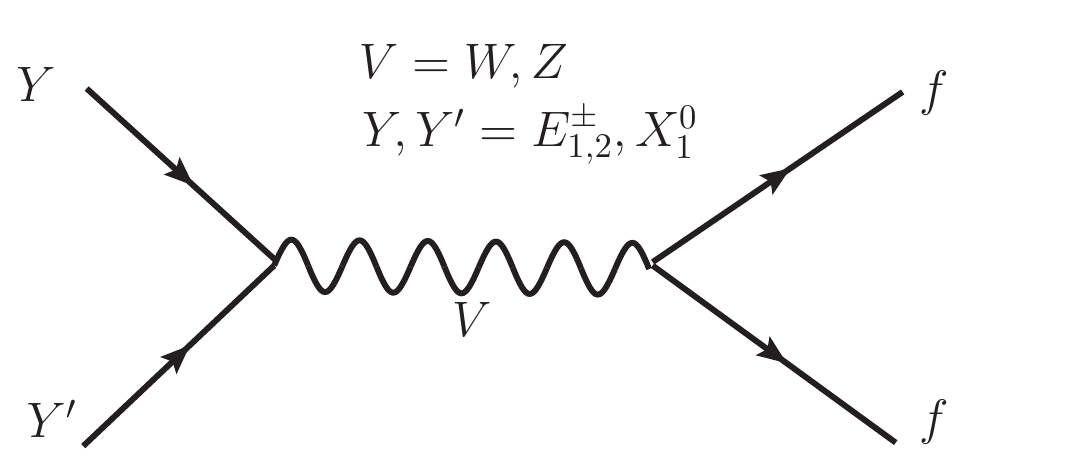}}
		\subfigure[]{
			\includegraphics[scale=0.6]{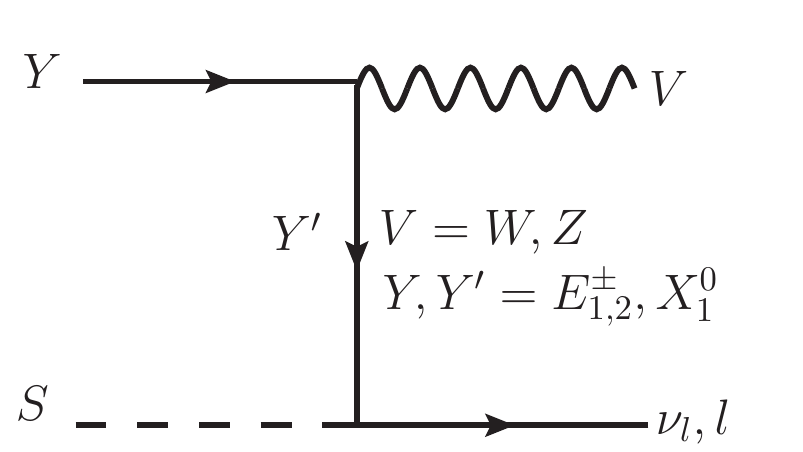}}
		\caption{ \it The co-annihilation and annihilation diagrams of the dark matter and the other $Z_2$-odd fermion fields. $f$ are standard model leptons and quarks.
		}
		\label{fig:DarkCoan}
	\end{center}
\end{figure}
\begin{figure}[h!]
	\begin{center}
		{\includegraphics[scale=.6]{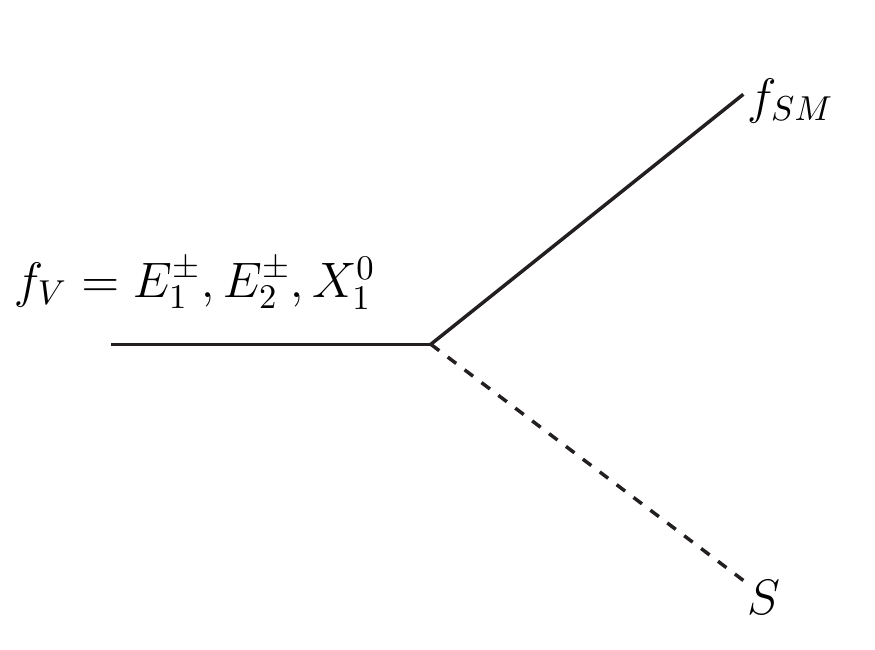}}
		{\includegraphics[scale=.6]{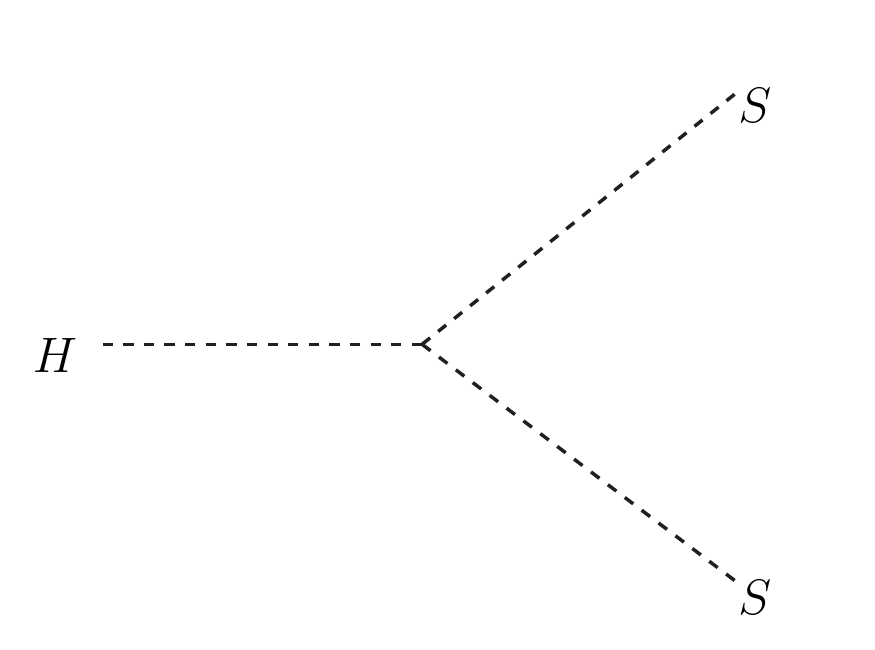}} 
		{\includegraphics[scale=.6]{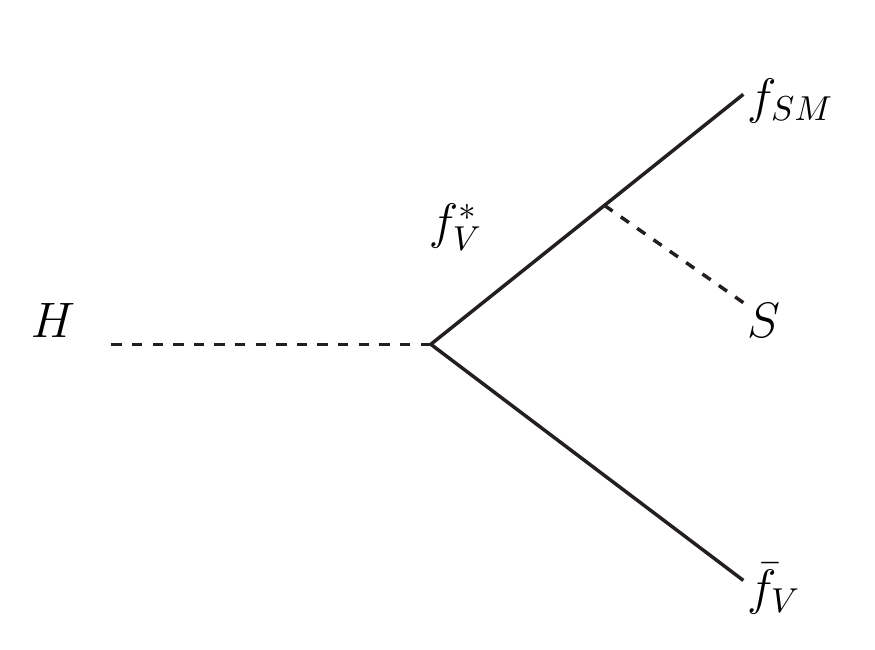}} 
		
		\caption{\it   \label{fig:decayHDD} \textit{ Dark matter production diagrams from the decay of the heavy particles contribute to the relic density.}}
	\end{center}
\end{figure}

where the Riemann zeta function has the value $\zeta_3=1.2$ and $g^*$ is the effective degrees of freedom in this framework. The thermally averaged annihilation cross-section of the particles can be expressed as~\cite{Gondolo:1990dk, Plehn:2017fdg},
\beq
<\sigma_{xx} v> = \frac{  2 \pi^2 T \, \int_{4 m^2}^\infty ds \sqrt{s} \, (s-4 m^2)  \, K_1(\frac{\sqrt{s}}{T})  \sigma_{xx}   }{   \left( 4 \pi m^2 T K_2(\frac{m}{T})  \right)^2        }.
\eeq
$K_{1,2}$ is the order 1 and 2 modified Bessel functions.
In this model, the dark matter production through annihilation depends upon the Higgs portal couplings $\kappa$ through $s$- and cross-channels (one can reverse the Figs.~\ref{fig:DarkAn}-(a), ~\ref{fig:DarkAn}-(b) and ~\ref{fig:DarkAn}-(c)) and the new Yukawa coupling $Y_{fi}$ through $t$-channel ( Figs.~\ref{fig:DarkAn}-(d)).
We checked~\cite{Das:2020hpd, Das:2021zea} that for $\kappa, Y_{fi} \sim \mathcal{O}(0.001)$ with dark matter mass $\sim \mathcal{O}$(GeV), the dark matter remains in thermal equilibrium at early Universe, $i.e.$, $\frac{ n_{eq}<\sigma_{xx} v>}{H(T)} >> 1$. These dark matter parameter spaces give the exact relic density via the Freeze-out mechanism. 
We also found~\cite{Das:2021zea} that the non-thermally produced dark matter (see Fig.~\ref{fig:decayHDD}) can also (depends on the parameters $\kappa$, $Y_{fi}$ and $\beta$) give the dark matter density in the right ballpark.
We now briefly discuss the dark matter regions coming from the Freeze-out and Freeze-in mechanisms in the following two subsections keeping an eye on all the constraints.

\vspace{-0.5cm}
\subsection{WIMP dark matter}
The singlet scalar $S$ as WIMP dark matter candidate gives exact relic density for the parameter spaces with mass $\mathcal{O}(100)$ GeV and annihilation coupling strength $\mathcal{O}(0.1)$~\cite{Das:2020hpd}. 
We use {\tt FeynRules}~\cite{Alloul:2013bka} to get the input codes for {\tt micrOMEGAs}~\cite{Belanger:2018mqt} and compute the relic density. We also verified these using {\tt SARAH-4.14.3}~\cite{Staub:2013tta,Staub:2015kfa} including {\tt SPheno-4.0.3}~\cite{Porod:2011nf} mass spectrum in {\tt micrOMEGAs}.
The Higgs portal couplings $\kappa$ controls the dark matter production and/or annihilation through $s$- and cross-channels (see figs.~\ref{fig:DarkAn}-(a), ~\ref{fig:DarkAn}-(b) and ~\ref{fig:DarkAn}-(c)).
The new Yukawa couplings $Y_{fi}$, $Y_{fi}^{\prime}$ and $Y_N$ also have huge influences~\cite{Das:2020hpd} to get the allowed dark matter parameter spaces. 
The main coupling vertices for the dark matter annihilation and co-annihilation can be written as~\cite{Das:2020hpd},
\begin{eqnarray}
g_{HSS}&=&  | \kappa v|,  ~g_{HHSS}=|\kappa|,~~g_{SiE_1^\pm}=A_{fi}=  |(\cos\beta Y_{fi} + \sin\beta Y_{fi}^{\prime})| ,\nn\\
&& g_{SiE_2^\pm}=B_{fi}= |(\sin\beta Y_{fi} + \cos\beta Y_{fi}^{\prime})|,\\
&& g_{S\nu_i X_1^0}=C_{fi}= | Y_{fi} , ~{\rm with}~i=e,\mu,\tau|. \nn
\label{eq:annivertex}
\end{eqnarray}
Here the interference between the $s$-, cross- and $t,u$-channels played a crucial role in achieving the exact dark matter density.  The co-annihilation channels (e.g., see Fig.~\ref{fig:DarkCoan}) also have an important role when the mass difference between the dark matter and other $Z_2$-odd particles is within $2\%-10\%$~\cite{Griest:1990kh}. For $\Delta M^{\pm,0}<0.1 M_{DM}$~\cite{Griest:1990kh} ($\Delta M^\pm= M_{E_1^\pm}-M_{DM}$ and $\Delta M^0= M_N-M_{DM}$), the co-annihilation channels (see Fig.~\ref{fig:DarkCoan} (b)) play an important role to get the dark matter density. It is to be noted that the contribution from the diagram corresponding to $S f_{V}\rightarrow f_{SM} H$, mediated by the t-channel exchange of a $f _V$ is very small due to propagator-vertex suppression. The co-annihilation effects go away for $\Delta M^{\pm,0}> 0.1 M_{DM}$.

\begin{figure}[h!]
	\centering
	\includegraphics[scale=.34]{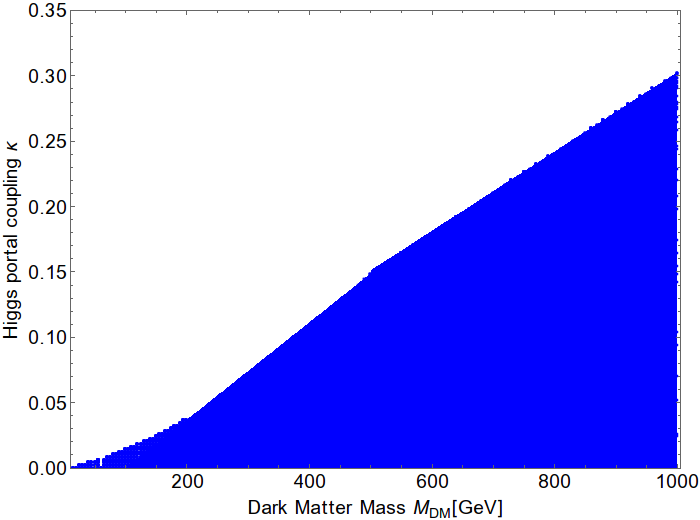}
	\includegraphics[scale=.34]{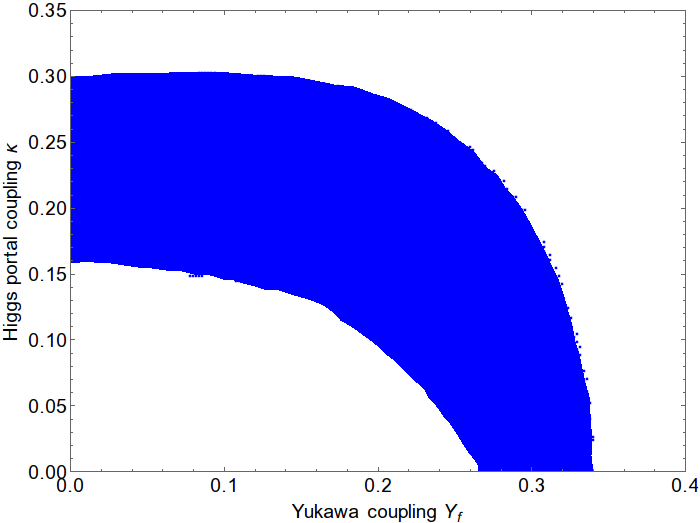}
	\caption{\it  The blue band indicates the relic density within the $3\sigma$ range. These plots are generated by varying the dark matter mass, Higgs portal coupling $\kappa$ and the new Yukawa coupling $Y_f$. The empty region are disfavoured by the relic density and direct detection bounds. The parameters allowed by relic density are also allowed by the recent LFV, electron as well as muon anomalous magnetic moment.}	\label{fig:relicall}
\end{figure}
 
We keep fixed the mass parameters $M_{E_1^{\pm}}=1500$ GeV and $M_{E_{2}^{\pm}}=3000$ GeV due to collider constraints as in~\cite{Das:2020hpd}. The effect of $- \, Y_{fi}^{\prime }\, \overline{l}_{i,R} E_S S$ is tiny for $ Y_{fi}^{\prime } =\mathcal{O}(0.1)$ as $M_{E_{2}^{\pm}}=3000$ GeV with $\cos\beta=0.995$.
For simplicity, we consider the effect from the interaction term $- \, Y_{fi}^{\prime }\, \overline{l}_{i,R} E_S S$  (see eqn.~\ref{lint}) to be zero, with $Y_{fi}^{\prime}=0$. 
The detailed can be found in Ref.~\cite{Das:2020hpd,Das:2021zea}.
We also assume $Y_{f1}=Y_{f3}=Y_f$ with $Y_{f2}=0.001$~\footnote{ Normally, the large Yukawa couplings are also allowed from the relic density limit. One can also get similar data points for the choice of $Y_{f2}=Y_{f3}=Y_f$ and $Y_{f1}=0.001$ or $Y_{f2}=Y_f$ with $Y_{f1}=Y_{f3}=0.001$. For this choice of couplings,  Fig. 4 (left) remains almost identical, whereas the right plot one gets changed due to having one dominant contribution from the diagram with the second Yukawa coupling only.}  to avoid flavor violating decays and fixed mixing angle as $\cos\beta=0.995$. This small mixing angle ($\beta$) diluted the contribution from the second charged fermion $E_2^\pm$.

\begin{table*}[h!]
	\centering
	\begin{tabular}{|p{1.6cm}|p{1.2cm}|p{1.1cm}|p{1.2cm}|p{1.2cm}|c|p{4.7cm}|}
		\hline
		\hline
		Channel & $M_{DM}$ (GeV) & ~~$\kappa~~~$& $M_{E_1^\pm}$ (GeV) &~~$Y_{f}$&$\Omega_{DM}h^2$&~~~~~~~~~~Percentage \\
		\hline

		&&&&&&$\sigma(S S\rightarrow \nu \nu)\quad~98\%$\\
			~~BP-1&8&0.00&1500~~&0.289&0.1198& $\sigma(SS \rightarrow  ll)\quad 2 \%$\\
		\hline
		&&&&&&$\sigma(S S\rightarrow \nu \nu)\quad~94\%$\\
		~~BP-2&80&0.01&1500~~&0.279&0.1157& $\sigma(SS \rightarrow  W^\pm W^{\mp*})\quad 6 \%$\\
		\hline				&&&&&&$\sigma(S S\rightarrow \nu\nu)\quad~74\%$\\
		&&&&&&$\sigma(S S\rightarrow W^{\pm}W^\mp)\quad~11\%$\\
		~~BP-3&250&0.035&1500~~&0.265&0.1228& $\sigma(SS \rightarrow  HH^*)\quad 6 \%$\\
		&&&&&&$\sigma(S S\rightarrow ZZ)\quad~5\%$\\
		&&&&&&$\sigma(S S\rightarrow t\bar{t}^*)\quad~3\%$\\
		\hline				&&&&&&$\sigma(S S\rightarrow \nu\nu)\quad~72\%$\\
		&&&&&&$\sigma(S S\rightarrow W^{\pm}W^\mp)\quad~13\%$\\
		~~BP-4&500&0.08&1500~~&0.280&0.1134& $\sigma(SS \rightarrow  HH^*)\quad 7 \%$\\
		&&&&&&$\sigma(S S\rightarrow ZZ)\quad~6\%$\\
		&&&&&&$\sigma(S S\rightarrow t\bar{t}^*)\quad~2\%$\\
		\hline
				&&&&&&$\sigma(SS\rightarrow W^\pm W^\mp)\quad 48\%$ \\
		~~BP-5&760&0.224&1500& 0.09 &0.1243& $\sigma(SS\rightarrow HH) \quad24\%$\\
				&&&&&&$\sigma(S S\rightarrow ZZ)\quad~24\%$\\
						&&&&&&$\sigma(S S\rightarrow ll)\quad~3\%$\\
		\hline
				&&&&&&$\sigma(SS\rightarrow W^\pm W^\mp)\quad 45\%$ \\
		~~BP-5&1000&0.285&1500& 0.185 &0.1230& $\sigma(SS\rightarrow HH) \quad 22\%$\\
				&&&&&&$\sigma(S S\rightarrow ZZ)\quad~22\%$\\
				&&&&&&$\sigma(S S\rightarrow \nu\nu)\quad~9\%$\\
						&&&&&&$\sigma(S S\rightarrow t\bar{t}^*)\quad~2\%$\\
		\hline
	\end{tabular}
	\caption{\it $\sigma(SS\rightarrow \nu \nu)$ is mainly dominated by the $t+u$-channel annihilation processes whereas $\sigma(SS\rightarrow YY),\,Y=W,Z,H,t$ dominated by the  $s+cross$-channel annihilation processes.}
	\label{tabDM:3}
\end{table*}

We now scan a prominent region for the dark matter mass in this work, Higgs portal coupling $\kappa$ and new Yukawa coupling $Y_f$.
The dark matter mass $M_{DM}$ has been scanned from $\sim 5$ GeV to 1000 GeV with a step size of $5$ GeV while Higgs portal coupling changes from $0$ to $0.35$ with step size $0.001$ and the new Yukawa coupling from $0$ to $0.5$ with step  $0.001$. 
We display the allowed parameters in $\kappa-M_{DM}$ plane in Fig.~\ref{fig:relicall}(left) and $\kappa-Y_f$ plane in Fig.~\ref{fig:relicall}(right).
In the low mass region (below $M_H/2$), $t,u$-channel annihilation processes play a key role in giving rise to correct relic bounds.
The Higgs portal coupling is kept very small ($\kappa\sim 0$) to avoid the Higgs signal strength constraints. Hence these regions are also allowed by Higgs decay width and direct detection bounds.
The above and below region of Fig.~\ref{fig:relicall}(left) correspond to Higgs portal coupling $|\kappa| \gtrsim \frac{M_{DM}}{3324.92~{\rm GeV}}$ are strictly ruled out from the direct detection data.
It is to be noted that for ($Y_f=0$), DM-mass region in between 70 GeV to 650 GeV is ruled out by the present direct detection cross-section~\cite{Aprile:2018dbl}.
One can adjust the new Yukawa coupling $Y_f$ to get the allowed region. 
We here choose a small Higgs portal coupling $\kappa$ and increase the Yukawa coupling $Y_f$ with dark matter mass to get the relic density at the right ballpark.
The combination of  $s$-,$t$-, and $u$- annihilation and co-annihilation channels can now give the exact dark matter density of the Universe. A similar effect can be seen in Fig.~\ref{fig:relicall}(right). We also show a few BMPs, including the percentage of the process channels in table.~\ref{tabDM:3}.

\subsection{FIMP dark matter}
This model can produce dark matter from the decay of the heavy fermions ($X_{Heavy}=E_1^\pm, E_2^\pm$ and $X_1^0$) and Higgs. It can also be produced from the annihilation of the bath particles. We find that all the mother particles (SM particles, including heavy fermions) remain in the thermal equilibrium in the early Universe as  ${\Gamma / H }>> 1$. We also check that the dark matter production from annihilation has a tiny contribution to dark matter relic density as compared to decay,~\cite{Das:2021zea}. 
The dark matter density ($M_{S}\equiv M_{DM}$) can be written as~\cite{Das:2021zea,Plehn:2017fdg,Hall:2009bx,Biswas:2016bfo},
\bea
\Omega h^2 = \frac{h^2}{3\, H_0^2\, M_{Pl}^2} \, \frac{ M_{ S }}{28} T_0^3 \, Y(x_0)\approx 1.09 \times 10^{27} \, M_{ S }  \,\sum_i  \frac{g_{X_{Heavy,i}} \, \Gamma (X_{Heavy,i} \rightarrow  f_{\rm SM}   S , S    S  ) }{ M_{X_{Heavy,i}}^2} 
\label{eq:totalOmega}
\eea
where, $X_{Heavy}=E_1^\pm,E_2^\pm, X_1^0, H$ and $f_{\rm SM} $ is standard model leptons. Hence, one need a very tiny decay partial decay width to get the dark matter density in the right ballpark.
\begin{figure}[h]
	\centering
	\includegraphics[scale=.350]{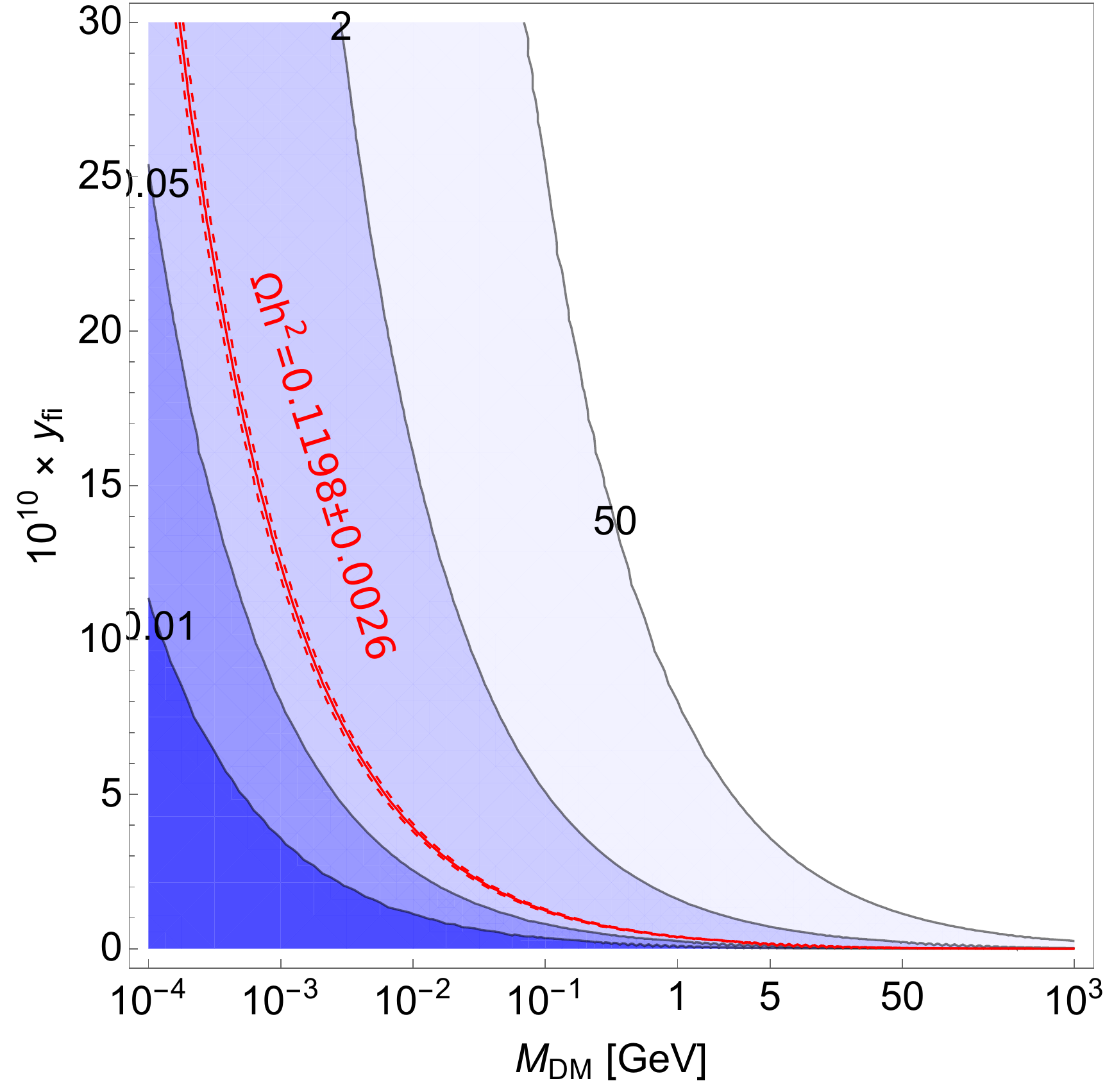}
	\includegraphics[scale=.3560]{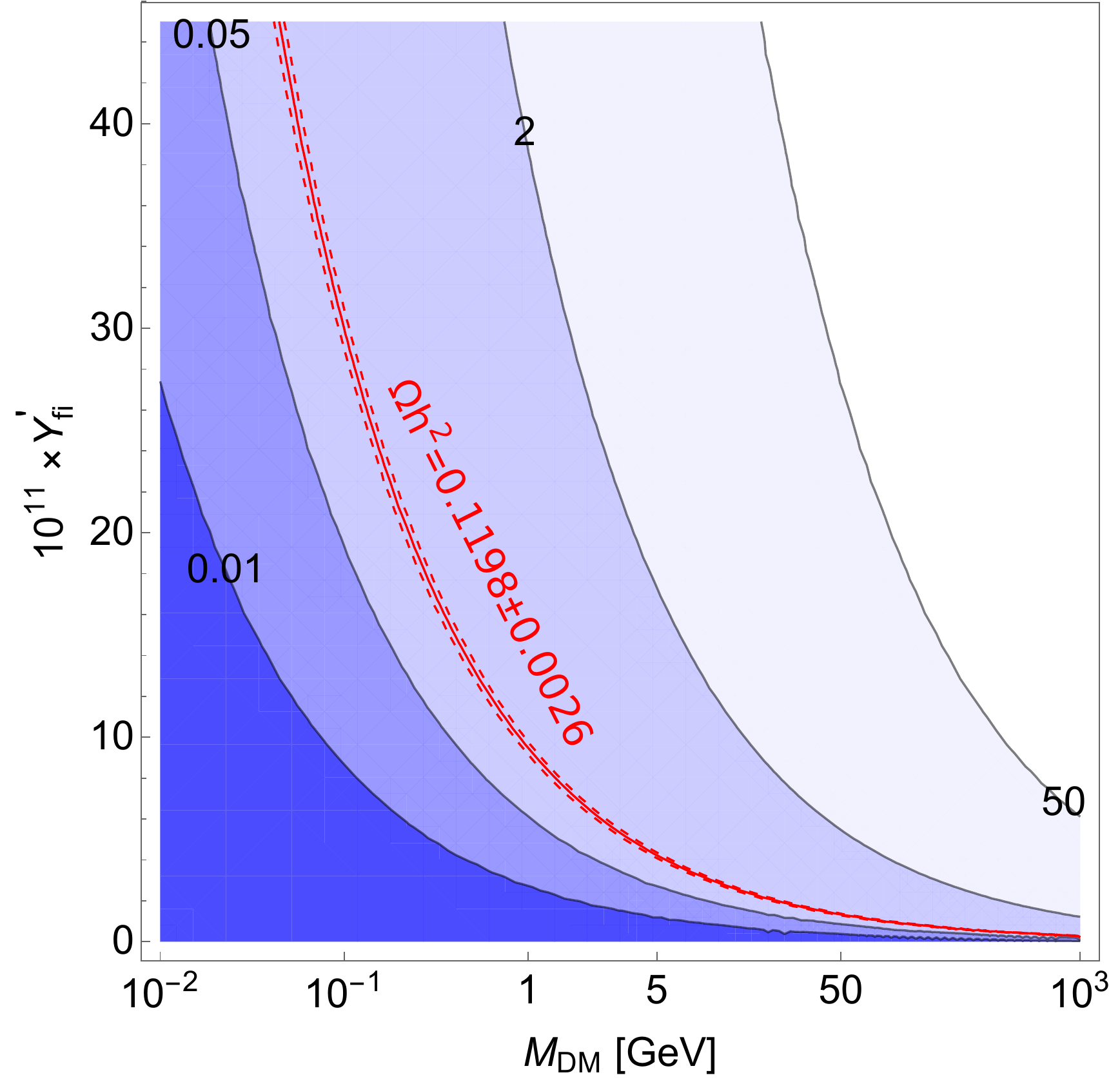}
	\caption{\it  The contour lines stand for the relic density in the new Yukawa coupling vs. dark matter mass plane. The red-lines indicate the relic density within the $3\sigma$ range. These region with $Y_{fi},Y_{fi}^{\prime }<<1$ are also allowed by the recent LFV, electron as well as muon anomalous magnetic moment limits.}\label{Fig:relicFIMP1}
\end{figure} 
The main production diagrams from the decay of the heavy particles are shown in Fig.~\ref{fig:decayHDD}.
The last diagram is kinematically forbidden as we have considered $M_{f_V} \sim 1.5$ TeV (LHC future search bound~\cite{Das:2020hpd}).

The partial decays of the heavy fermions and Higgs into the dark matter particle are  given by,
\bea
\Gamma(f_V \rightarrow  f_{\rm SM}   S ) = \frac{M_{f_V}}{8 \pi } \, |g_{f_V \, f_{SM} \,   S }|^2 ~\&~ \Gamma(H \rightarrow   S    S ) = \frac{1}{32 \pi \,M_{H}} \, |g_{H  S    S }|^2 \, \left( 1 -  \frac{M_{ S }^2}{M_{H}^2}  \right)^{1 \over 2},
\label{eq:decayMF}
\eea
where, the coupling strengths are (see eqn.~\ref{eq:annivertex}) $
g_{HSS}= | \kappa v|$, $g_{SiE_1^\pm}= A_{fi}=  |(\cos\beta Y_{fi} + \sin\beta Y_{fi}^{\prime})|$, $g_{SiE_2^\pm}= B_{fi}= |(\sin\beta Y_{fi} + \cos\beta Y_{fi}^{\prime})|$ and $
g_{S\nu_i X_1^0}=C_{fi}= | Y_{fi}| , ~{\rm with}~i=e,\mu,\tau$.

\begin{figure}[h]
\begin{center}
	\includegraphics[scale=.350]{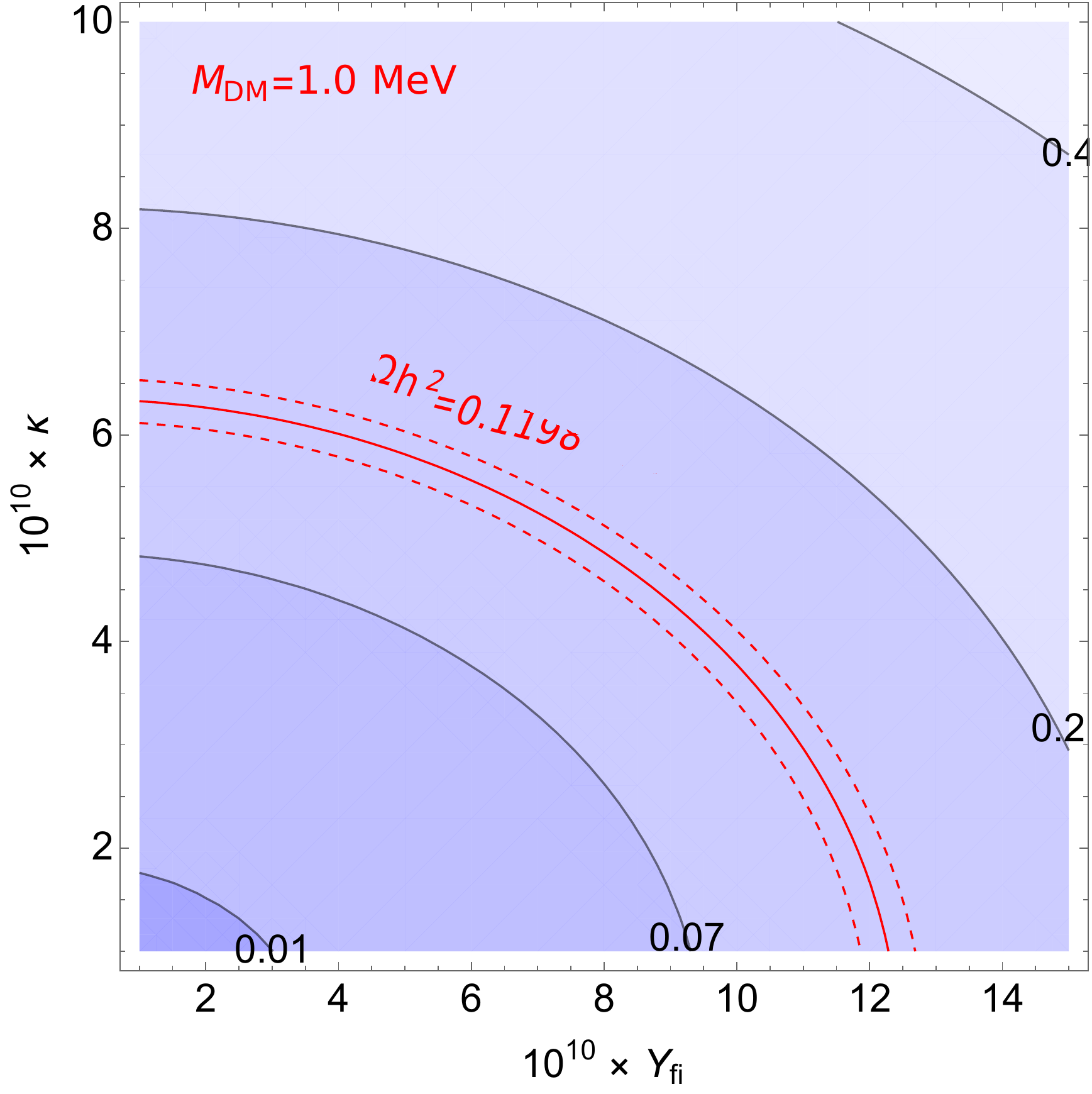}
	\includegraphics[scale=.3560]{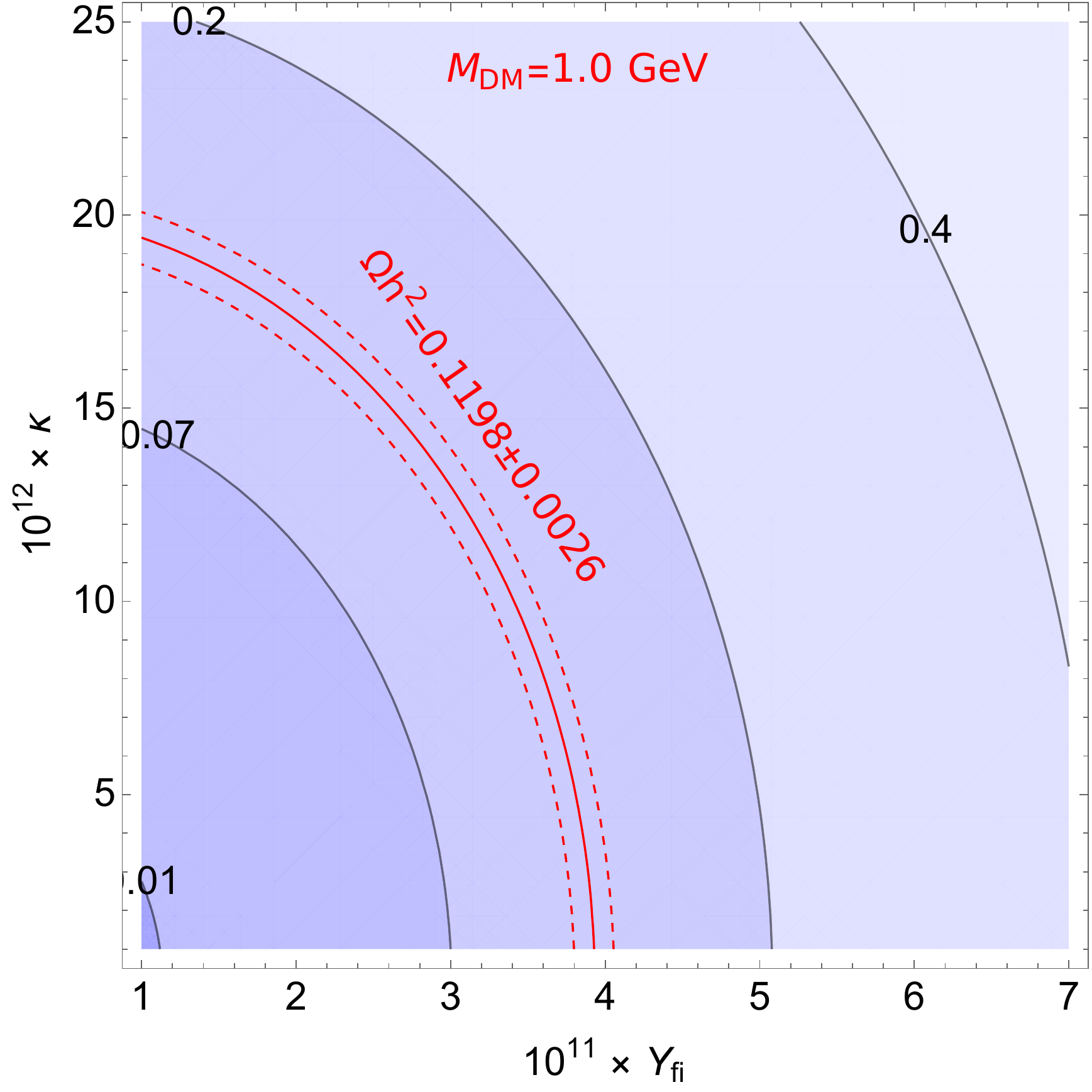}
	\caption{\it  The color coding same as in Fig.~\ref{Fig:relicFIMP1}. We neglect the contributions from term $- \, Y_{fi}^{\prime }\, \overline{l}_{i,R} E_S S$.}\label{Fig:relicFIMP2}
\end{center}
\end{figure}

\begin{figure}[h]
\begin{center}
	\includegraphics[scale=.350]{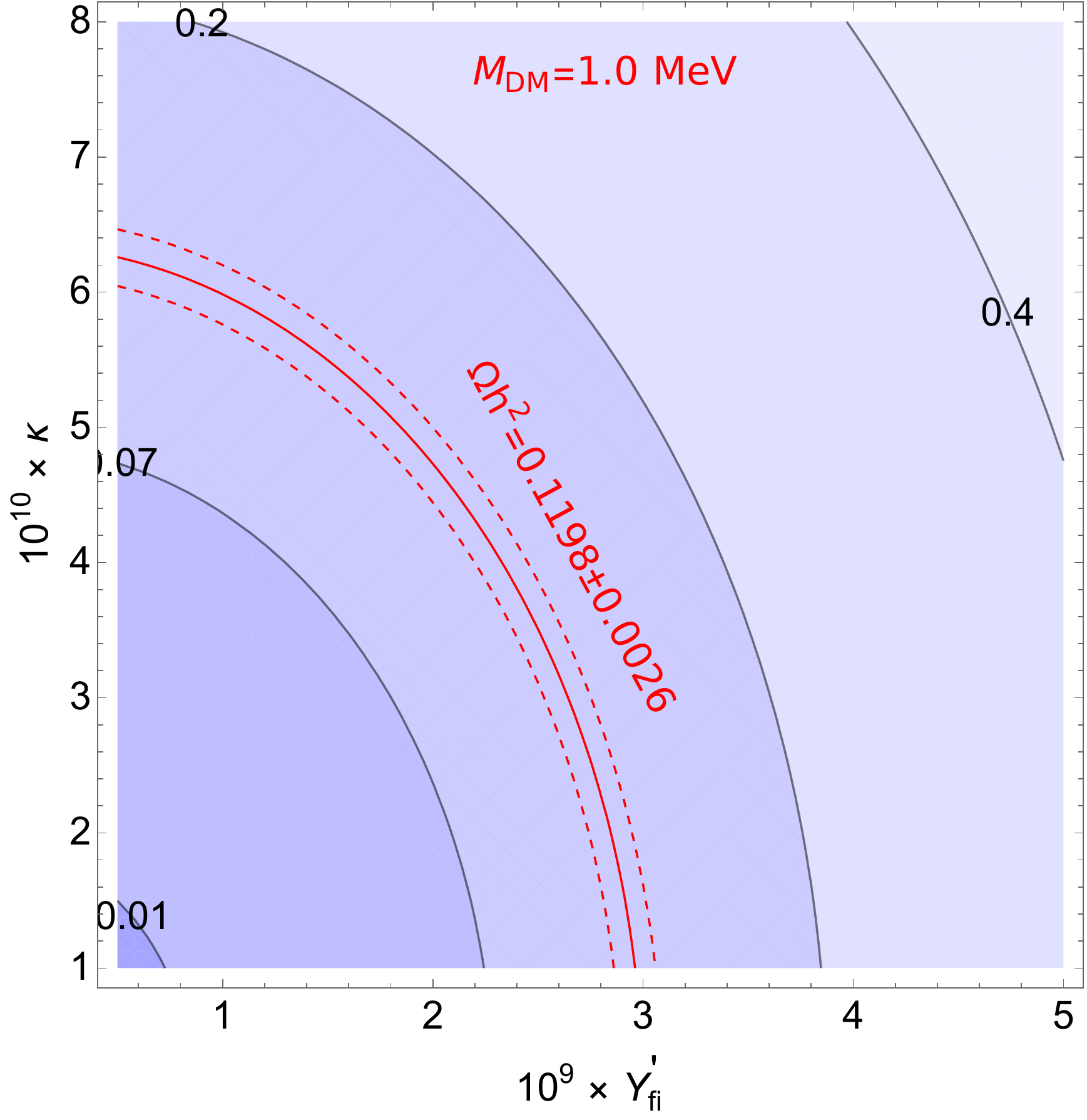}
	\includegraphics[scale=.3560]{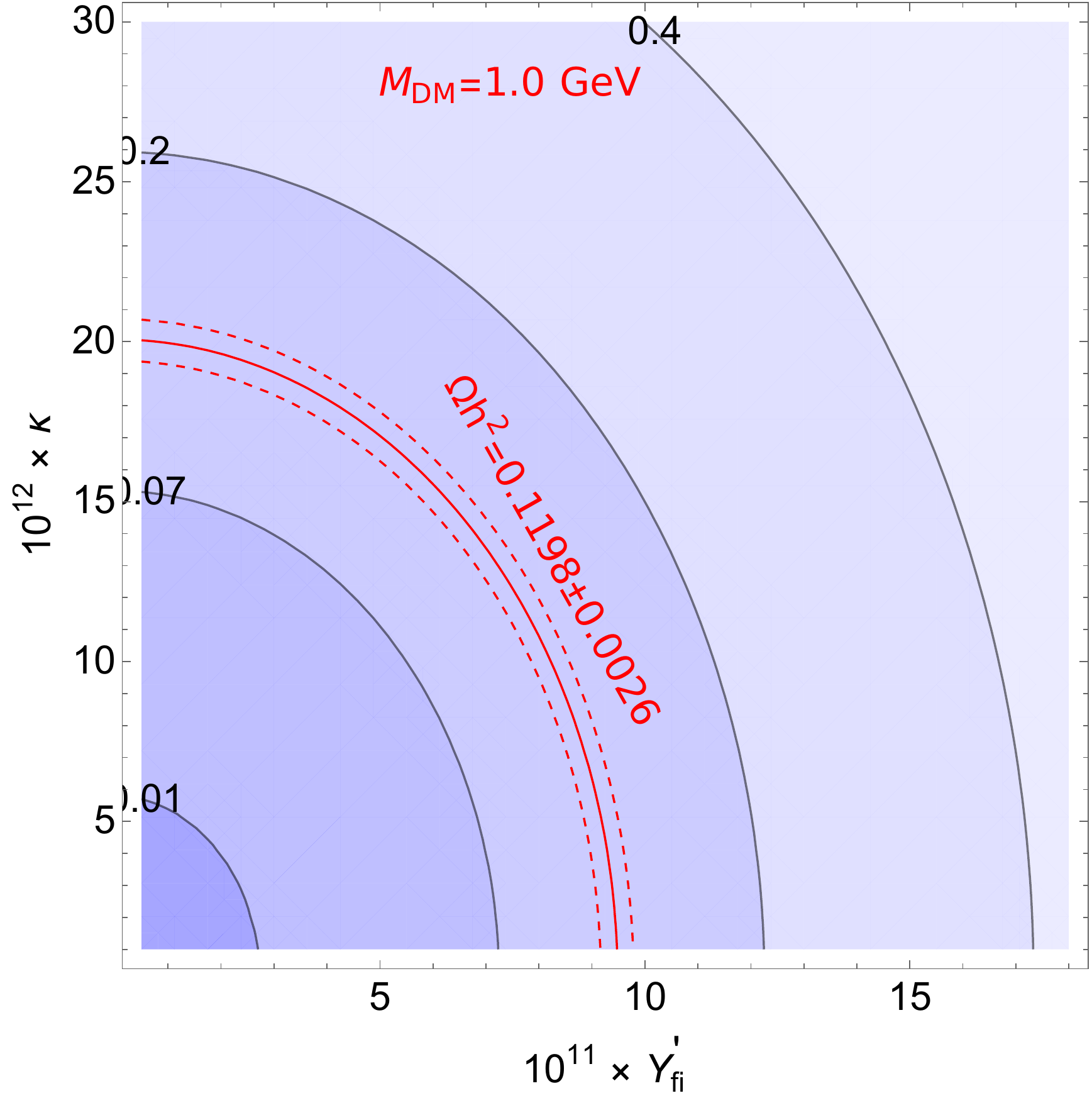}
	\caption{\it  The color coding same as in Fig.~\ref{Fig:relicFIMP1}. We neglect the contributions from term $- \, Y_{fi} \overline{\psi}_{i,L} F_D S$.}\label{Fig:relicFIMP3}
\end{center}
\end{figure}

Let us first neglect the contribution from the decay and annihilation from the scalar sector, i.e., $\kappa=0$.
We show the variation of the new vector-like Yukawa coupling $Y_{fi}$ ($i=e,\mu,\tau$) with the mass of the dark matter in Fig.~\ref{Fig:relicFIMP1} (left). Here we also neglected the effect from the interaction term $- \, Y_{fi}^{\prime }\, \overline{l}_{i,R} E_S S$ by considering $Y_{fi}^{\prime}=0$ in the left plot. In the right plot, we now ignore the contributions from term $- \, Y_{fi} \overline{\psi}_{i,L} F_D S$.
The solid red line represents $\Omega h^2 = 0.1198$, and the red dashed lines correspond to the $3\sigma$ variation in $\Omega h^2$. The lighter region corresponds to higher values of $\Omega h^2$, which over close the Universe, and these regions are strictly forbidden. One can see that we need $Y_{fi}, Y_{fi}^{\prime} \lesssim \mathcal{O}(10^{-10})$ to have $\Omega h^2 = 0.1198$ for $1$ MeV dark matter.
The combined effect of the Higgs portal coupling and vector-like Yukawa couplings are shown in Figs.~\ref{Fig:relicFIMP2} and \ref{Fig:relicFIMP3} for two different sets of dark matter masses. We also see that to get a viable FIMP; one needs a tiny Higgs portal coupling along with the Yukawa couplings. The Universe will overclose for the large value of these couplings.
\vspace{-0.5cm}
\section{Stability Analysis}
\label{sec:stab}
We again remind the readers that depending on the additional coupling strength(s), the new scalar pulls the electroweak vacuum towards stability, whereas the fermion pushes it towards instability~\cite{Garg:2017iva, Goswami:2018jar, Khan:2015ipa, Khan:2014kba}.
Thus, in this radiative seesaw model, the Higgs portal and new Yukawa couplings, as well as the new field masses, will play a crucial role in the stability/metastability of the electroweak vacuum. We here discuss how the stability of the electroweak vacuum is modified in the presence of the extra fermionic/scalar couplings if we assume that there is no other new physics up to the Planck mass scale $M_{Pl}$. 
We also discuss the proper matching conditions. It will give the values of the model parameters at the electroweak scale. Using them into the renormalizable group equations, calculate these parameters at a high scale up to the Planck scale. We use proper evolution of these couplings and calculate the tunneling of the electroweak vacuum and  put stringent bounds on the model parameters from the instability, metastability,
stability, and perturbative-unitary.
In the following, we now discuss the detailed theoretical background and tools needed in the stability analysis of the electroweak vacuum up to the Planck scale. 

The standard model one-loop effective Higgs potential in the $\overline{MS}$ scheme and the Landau-gauge is given by,
\begin{equation}
V_1^{\rm SM}(h)=\sum_{i=1}^5 \frac{n_i}{64 \pi^2} M_i^4(h) \left[ \ln\frac{M_i^2(h)}{\mu^2(t)}-c_i\right] \,,
\end{equation}
here the index $i$ is summed over all the standard model fields,  $M_i^2(h)= \kappa_i(t)\, h^2(t)-\kappa_i^{\prime}(t)$ and $c_{W,Z}=5/6$, $c_{h,G,f}=3/2$ ~\cite{Casas:1994qy, Altarelli:1994rb, Casas:1994us, Casas:1996aq, Quiros:1997vk}.  The number of degrees of freedom of the particle fields is denoted by $n_i$.  The values of  $n_i$, $\kappa_i$ and $\kappa_i'$ for the standard model are given in the eqn.(4) in \cite{Casas:1994qy}. The above contribution comes with a positive sign for the scalar and gauge bosons, whereas it is negative for the fermion fields. The running energy scale $\mu$ is related to a dimensionless parameter $t$ as $\mu(t)=M_Z \exp(t)$. The extra contribution to the one-loop effective Higgs potential from the new fermion and scalar can be written as,
\bea
 V_1^{f_{Heavy}+s} (h) &=& - \frac{3 (M_{E_{Heavy}}^\dag (h) M_{E_{Heavy}}(h))^2}{64 \pi^2} \Bigg[ \textrm{ln}\,\frac{M_{E_{Heavy}}^\dag (h)M_{E_{Heavy}}(h)}{\mu^2(t)} -\frac{3}{2}\Bigg]  \nn\\
 &&+ \frac{M_{S}^4}{64 \pi^2} \Bigg[ \textrm{ln}\,\frac{M_{S}(h)^2}{\mu^2(t)}  -\frac{3}{2}\Bigg],
 \eea
where $M_{E_{Heavy}}(h) \equiv \frac{Y_N}{\sqrt{2}}\,h$ and $M_{S}\equiv M_{DM}$. We also change the singlet scalar field notation from $S$ to $s$ to avoid confusion with action of the scalar potential. In this work, we use the two-loop contributions to the effective potential from the standard model fields whereas one-loop from the extra fermion and scalar part.
For very high field value $h(t) \, >> \, v_{SM}$, the effective potential can be approximated as,
 $ V_{eff}^{SM+f_{Heavy}+s}\, = \, \lambda_{eff,h}(h) \frac{h^4}{4} $. The standard model expressions at one- and two- loop  for $\lambda_{eff}(h)$ can be found in Ref.~\cite{Buttazzo:2013uya}.
The contribution due to the extra fields are simplified as,
\be 
\lambda^{f_{Heavy}+s}_{eff,h}(h) \, = \, -  \frac{e^{4 \Gamma (h)}}{32 \pi^2} 3\, \Bigg( (Y_N^\dag Y_N)^2  \Big(   \textrm{ln}\,\frac{(Y_N^\dag Y_N)}{2} \,-\frac{3}{2} \Big) - \kappa^4  \Big(   \textrm{ln}\,\frac{\kappa(h)}{2} \,-\frac{3}{2}\ \Big) \Bigg).
\label{labmbdaeffh}
\ee
The factor $\Gamma(h) \, = \, \int_{M_t}^{h} \gamma_h(\mu)\, d\,\textrm{ln}\,\mu $
indicates the Higgs field wave function renormalization.
Here $\gamma_h(\mu)$ is the anomalous dimension of the Higgs~\cite{Casas:1994qy, Altarelli:1994rb, Casas:1994us, Casas:1996aq, Quiros:1997vk}. The additional contributions from these new particles are provided in Appendix~\ref{sec:app}. The contribution to which from the heavy-fermion and scalar at one loop is $\frac{3}{2} \mbox{Tr}\Big({Y_N  Y_N^{\dagger}}\Big)$ and $\kappa^2$. All the couplings are evaluated at the scale $\mu=h$. In this choice, all the running coupling constants ensure faster convergence of the perturbation series of the potential~\cite{Ford:1992mv}.

The contribution to the one-loop effective singlet scalar field ($s$) potential from the new fermion and scalar is given by,
\bea
 V_{s,1}^{f_{Heavy}+s+h} (s) &=& - \frac{2 (M_{F_{D}}^\dag (s) M_{F_{D}}(s))^2}{64 \pi^2} \Bigg[ \textrm{ln}\,\frac{M_{F_{D}}^\dag (s)M_{F_{D}}(s)}{\mu^2(t)} -\frac{3}{2}\Bigg]  \nn\\
 &&- \frac{ (M_{E_{S}}^\dag (s) M_{E_{S}}(s))^2}{64 \pi^2} \Bigg[ \textrm{ln}\,\frac{M_{E_{S}}^\dag (s)M_{E_{S}}(s)}{\mu^2(t)} -\frac{3}{2}\Bigg] \nn\\
 &&+ \frac{M_{S}^4}{64 \pi^2} \Bigg[ \textrm{ln}\,\frac{M_{S}(s)^2}{\mu^2(t)}  -\frac{3}{2}\Bigg],
 \eea
where $M_{E_{D}}(s) \equiv Y_{f}\,s$ and $M_{E_{S}}(s) \equiv Y_{f}^{\prime}\,s$.
Similarly, the effective potential along singlets scalar field direction can be approximated as, $ 
V_{eff}^{s}\, = \, \lambda_{eff,s}(s) \frac{s^4}{4 !}
$. The  expression for $ \lambda_{eff,s}(s)$ including one-loop can be simplified as,
\bea 
\lambda_{eff,s}(s) & =& e^{4 \Gamma_s (s)}  \Bigg( \lambda_{S}(s) - 2\, \frac{(Y_f^\dag Y_f)^2}{64 \pi^2}  \Big(   \textrm{ln}\,\frac{(Y_f^\dag Y_f)}{2} \,-\frac{3}{2} \Big) \nn\\
 && \quad \quad \quad - \frac{(Y_f^{\prime\, \dag} Y_f^{\prime})^2}{64 \pi^2}  \Big(   \textrm{ln}\,\frac{(Y_f^{\prime\, \dag} Y_f^{\prime})}{2} \,-\frac{3}{2} \Big)+ \frac{\kappa^4}{64 \pi^2}  \Big(   \textrm{ln}\,\frac{\kappa}{2} \,-\frac{3}{2}\ \Big) \Bigg).
\label{labmbdaeffS}
\eea
The couplings are evaluated at the scale $\mu=s$. The similar factor $\Gamma_s(s) \, = \, \int_{M_t}^{s} \gamma_s(\mu)\, d\,\textrm{ln}\,\mu $ indicates the singlet scalar field wave function renormalization. The function $\gamma_s(\mu)$ is given in Appendix~\ref{sec:app}. The wave function renormalization also plays an inportant role in the tunneling probability of the electroweak vacuum. Without this function, the lifetime could vary from $10^{5}$ to $10^{100}$ seconds, depending on the parameters. One must be very careful in handling this function in addition to the other renormalization group equations (RGEs).

We do the RG evolution up to the Planck scale of all the couplings to analyze the scalar potential along the Higgs and singlet scalar field directions. We first follow the threshold corrections as in Refs.~\cite{Sirlin:1985ux,Bezrukov:2012sa,Degrassi:2012ry,Khan:2014kba} to calculate all the standard model couplings at the top mass scale $M_t$.
We use one-loop RGEs for the $SU(2)$ and $U(1)$ gauge couplings $g_2(M_t)$ and $g_1(M_t)$
to calculate. We also check that the two-loop RGEs for $g_1$ and $g_2$ will do not change our result significantly.
For the $SU(3)$ gauge coupling $g_3(M_t)$, which is crucial in this analysis, we use three-loop RGEs. We consider the contributions from the five quarks and the effect of the sixth too, i.e., the top quark has been taken using an effective field theory approach. The detailed can be found in Refs.~\cite{Sirlin:1985ux,Bezrukov:2012sa,Degrassi:2012ry}.
We also take the leading term in the four-loop RGE for $\alpha_s$.
The proper matching between the top pole mass and the $\overline{MS}$ renormalized coupling is obtained by using the threshold correction  $ y_t(M_t) \, = \, \frac{\sqrt{2}M_t}{v}\,(1\,+\,\delta_t(M_t)) $, here $\delta_t(M_t)$ denotes the matching correction for $y_t$ at the top pole mass.
We similarly compute the Higgs quartic coupling $\lambda$ using $\lambda(M_t) \, = \, \frac{M_H^2}{2 v^2} \, (1\,+\,\delta_H(M_t)) $ relation.
The three loops QCD corrections~\cite{Melnikov:2000qh} have included to calculate this at the scale $M_t$. We also added the electroweak corrections up to one-loop \cite{Hempfling:1994ar,Schrempp:1996fb} and the $O(\alpha \alpha_s)$ corrections to the matching of top Yukawa and top pole mass \cite{Bezrukov:2012sa,Jegerlehner:2003py}.
By using these threshold corrections, we verified the standard model couplings at $M_t$ as in Refs~\cite{Buttazzo:2013uya, Khan:2014kba}.

We run these standard model couplings up to the heavy scalar and fermionic mass scale using the standard model RGEs~\cite{Chetyrkin:2012rz, Zoller:2012cv, Chetyrkin:2013wya, Zoller:2013mra} only.
The extra contributions due to the singlet scalar and fermionic fields (depending on the mass hierarchy) are included after the threshold heavy scalar and fermionic mass scale.
In this work, we consider the new fermion is heavier than the singlet scalar fields, hence the fermionic contributions added in the last at the threshold fermionic mass scale.
Then we evolve all the couplings up to the Planck scale $\mpl$ to find the position and depth of the new minima at the high scale along the Higgs and singlet scalar field directions. To calculate the decay probability of the electroweak vacuum to the true (deeper than EW) vacuum at the present epoch, we first have to minimize the effective action $S=\int	\mathcal{L} d^4x$ of the scalar potential.
One can use the bounce solution of the euclidean equations of motion of the scalar field~\cite{Coleman:1977py, Isidori:2001bm, Buttazzo:2013uya}.
The decay probability of the electroweak vacuum to the true vacuum is given by,

\be
{\cal P}_0=0.15 \frac{\Lambda_B^4}{H^4} e^{-S(\Lambda_B)},
\label{eq:probaction}
 \ee
where, $H$ stands for the Hubble constant and $S(\Lambda_B)$ is the minima action of the scalar potential at the bounce size $R=\Lambda_B^{-1}$. Conversely, $\Lambda_B$ is the running energy scale where effective action becomes minima.
For the $V(\phi)\approx \frac{1}{4} \lambda_{\phi} \phi^4$ ($\phi$ is any scalar),  the minima action can be approximated as $\,\,{\rm where}\,\,S(\Lambda_B)\approx \frac{8\pi^2}{3|\lambda_\phi (\Lambda_B)|}\,$~\cite{Coleman:1977py, Isidori:2001bm, Buttazzo:2013uya}.
The decay probability ${\cal P}_0 < 1$ implies that electroweak vacuum is in metastable, i.e., the lifetime of the Universe $\tau_U\sim 10^{17}$ secs~\cite{Ade:2015xua}. It can be translated into a bound on the quartic coupling $\lambda_{\phi}$,
\be 
\lambda_{ \phi} > \lambda_{ \phi~ \rm min}(\Lambda_B) = \frac{-0.063793}{1-0.00969537 \, \ln\left( {\Lambda_{r}}/{\Lambda_B} \right)} ,
\ee
\label{eq:eflammeta}
where, we consider a low energy reference point $\Lambda_{r}=1$ TeV. It will not affect our results as the decimal number in the above equation~\ref{eq:eflammeta} will also change with the chosen value of $\Lambda_{r}$. For the $V(\phi)\approx \frac{1}{24} \lambda_{\phi} \phi^4$, we can have have $\lambda_{ \phi} > \lambda_{ \phi~ \rm min}(\Lambda_B) = \frac{-6\times 0.063793}{1-0.00969537 \, \ln\left( {\Lambda_{r}}/{\Lambda_B} \right)} $. It is to be noted that one has to take RGEs carefully to compute the exact value of $\lambda_{ \phi~ \rm min}(\Lambda_B) $  at the bounce scale. 
 $\lambda_{\phi}(\Lambda_B)< \lambda_{\phi~ \rm min}(\Lambda_B)$ corresponds to the unstable region and the electroweak vacuum is absolutely stable at $\lambda_{\phi}(\Lambda_B) > 0$. In this model, the theory violates the perturbative unitarity at $\lambda_{eff,h}(\Lambda_B) > \frac{4\pi}{3}$ and $\lambda_{eff,s}(\Lambda_B) > 8\pi$ (see eqns.~\ref{eq:pert} and \ref{eq:unitary} for details).

So far, we have discussed that if there is a high scale minima (deeper than electroweak minima) solely along the Higgs $h$ or singlet scalar $s$ field directions. We also consider one of them above the electroweak minima or no minima in that direction. For this choice, we use these formulas to calculate the tunneling time. Suppose there exist two deeper minima than electroweak minima at a very high scale, both of which are far from each other. In this case, one can also use these formulae, but tunneling always happens into the deeper minima only.
If there exist a single minima between the $h$ and $s$ direction, then we have to consider the effect of the mixing term in the potential $ V(mix)= \frac{1}{4}\left\lbrace\kappa +   \sqrt{\frac{2 \lambda \lambda_S}{3}}\right\rbrace h^2 s^2$. If there is no minima exist along the $h$ and $s$ directions, then one has to carefully analyses this term from the scalar potential. It is to be noted that we do not find any allowed parameters in this model corresponding to a minima between the $h$ and $s$ directions.

\subsection{Running RGEs of the effective scalar quartic couplings}

It is now well understood that the electroweak vacuum of the Higgs potential is not the global minima. A quantum tunneling from the electroweak minima to the true (deeper) minima may occur. This tunneling happens because the RG running can make the effective quartic coupling $\lambda_{h, eff}$ negative at a high energy scale, and the behavior of the slope, i.e., $\beta_\lambda$ such that the potential form a deeper minima. In this model, we also find that the effective singlet quartic coupling $\lambda_{s, eff}$ becomes negative, and $\beta_{\lambda_S}$ changes in such a way that one can have additional minima in the singlet scalar field direction.

\begin{figure}[h!]
	\centering
	\includegraphics[scale=.34]{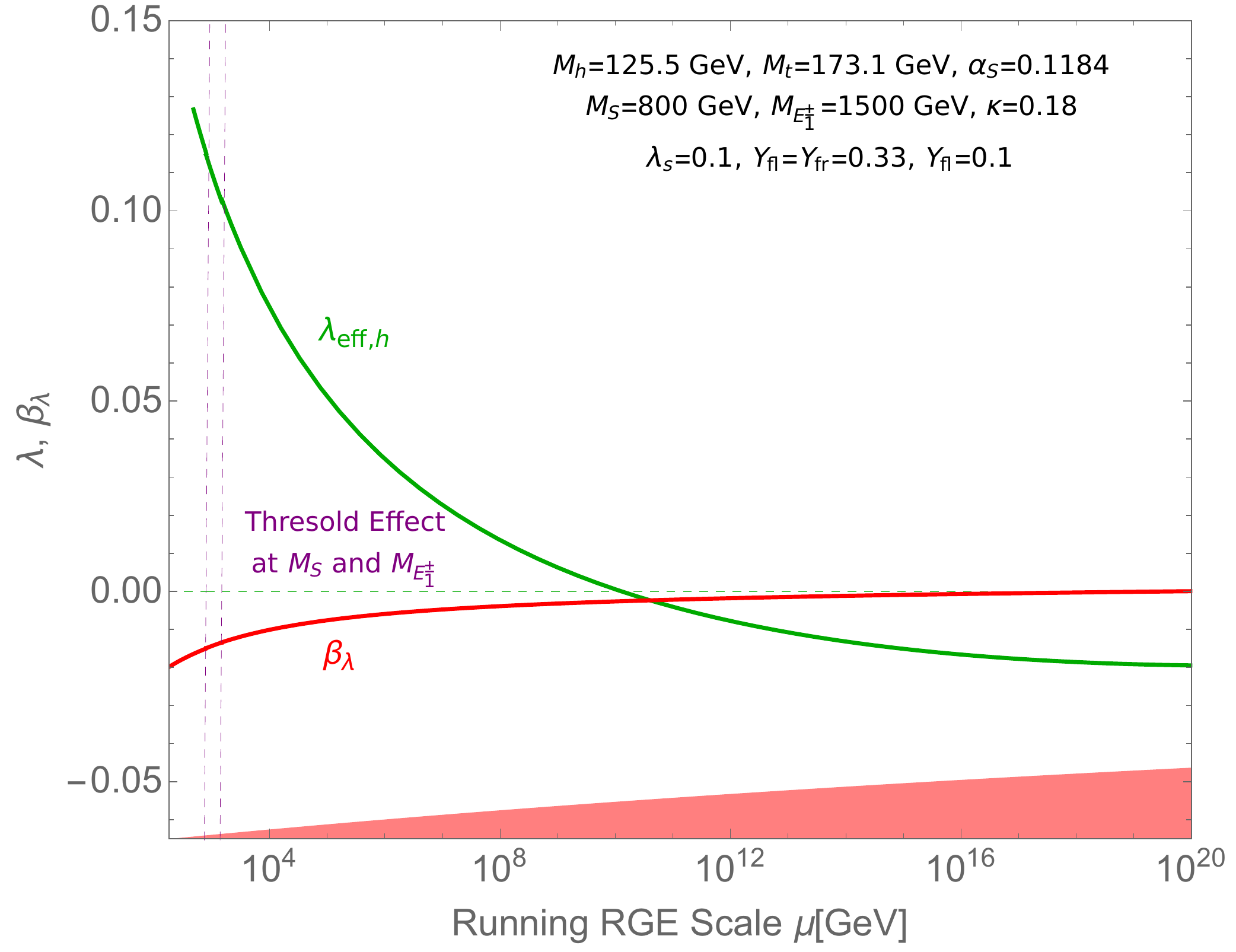}
	\includegraphics[scale=.34]{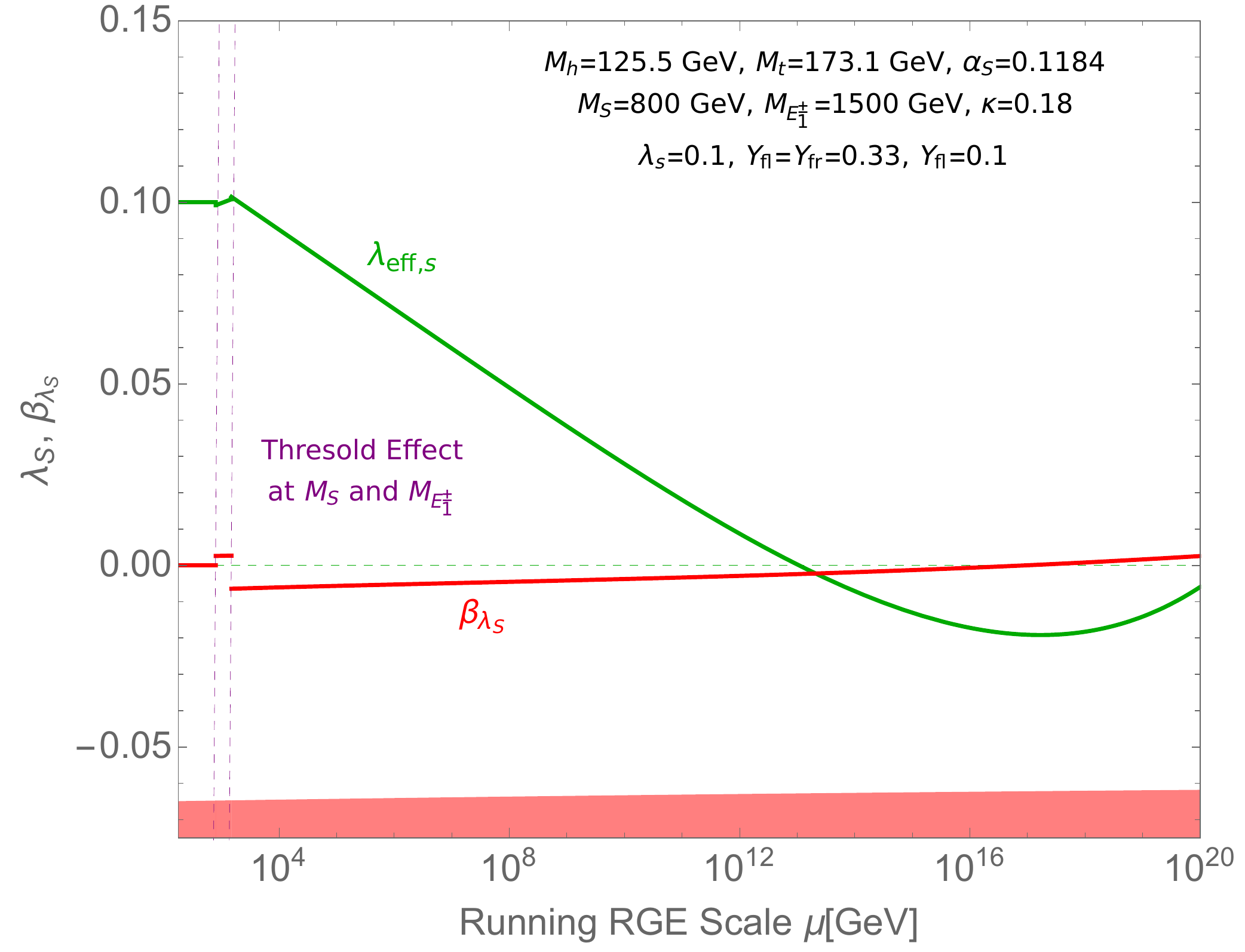}
	\caption{\it  RG evolution of the Higgs and singlet scalar quartic couplings. These plots left side for the running of $\lambda$ whereas right side for $\lambda_S$ with the choice of parameters $M_h=125.5$ GeV, $M_t=173.1$ GeV, $\alpha_S=0.1184$ and $M_{S}=800$ GeV, $M_{E_1^\pm}=1500$ GeV, $M_{E_2^\pm}=3000$ GeV with $\kappa=0.18$, $Y_{f}=Y_{f}^\prime$=0.33 and $Y_{N}=0.1$.}	\label{fig:s1}
\end{figure}

\begin{figure}[h!]
	\centering
	\includegraphics[scale=.34]{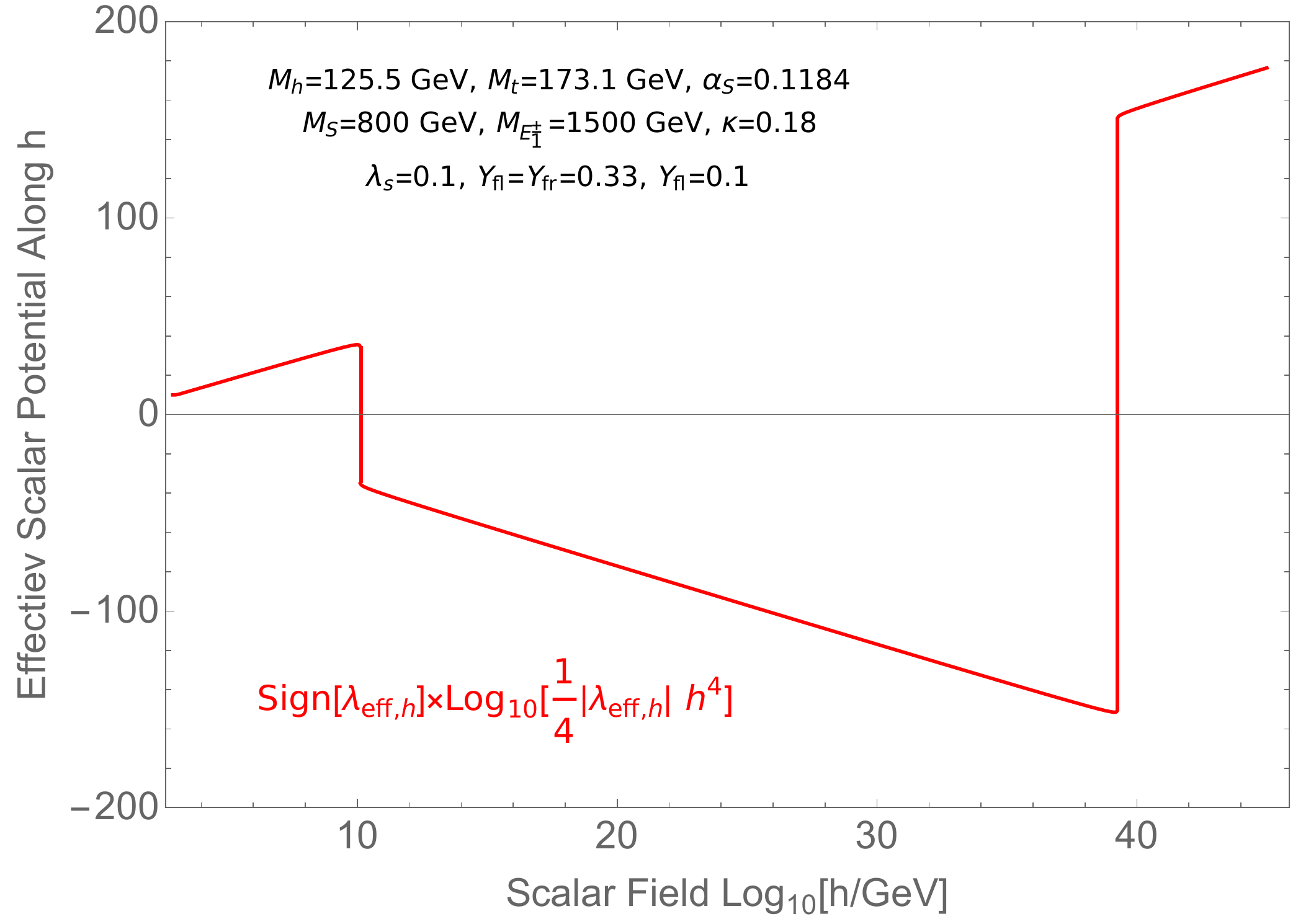}
	\includegraphics[scale=.34]{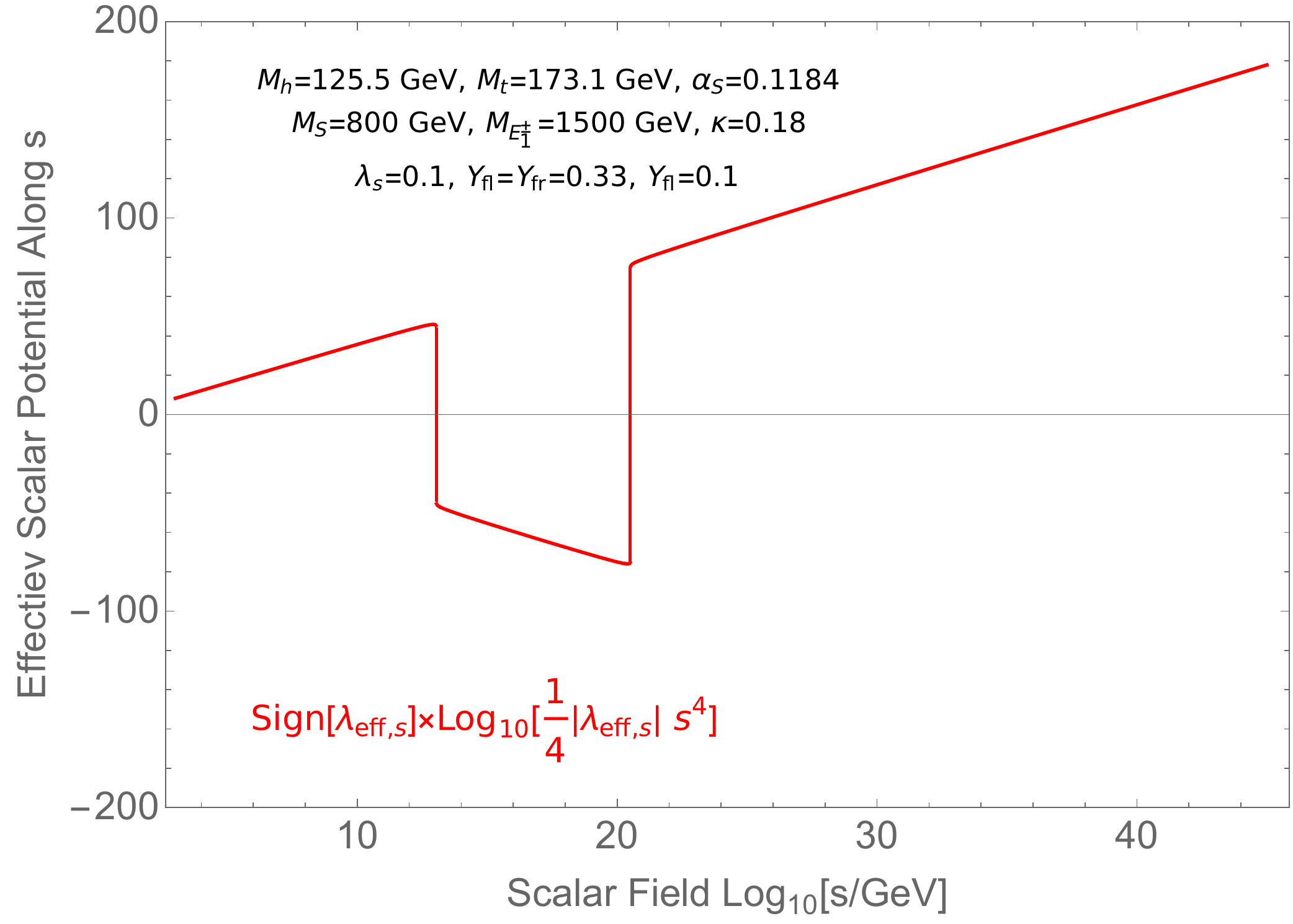}
	\caption{\it The effective potential for the choice of parameters $M_h=125.5$ GeV, $M_t=173.1$ GeV, $\alpha_S=0.1184$ and $M_{S}=800$ GeV, $M_{E_1^\pm}=1500$ GeV, $M_{E_2^\pm}=3000$ GeV with $\kappa=0.18$, $Y_{f}=Y_{f}^\prime$=0.33 and $Y_{N}=0.1$.}	\label{fig:s1pot}
\end{figure}

In Fig.\ref{fig:s1}, we show the running of the Higgs quartic and singlet scalar couplings  for our benchmark point $M_h=125.5$ GeV, $M_t=173.1$ GeV, $\alpha_S=0.1184$ and $M_{S}=800$ GeV, $M_{E_1^\pm}=1500$ GeV, $M_{E_2^\pm}=3000$ GeV with $\kappa=0.18$, $Y_{f}=Y_{f}^\prime$=0.33 and $Y_{N}=0.1$. This point gives dark matter density at $\Omega h^2=0.1198$ through freeze-out mechanism. We also keep fixed the central value of the standard model parameters $G_F=1.16637 \times 10^{-5}$ GeV$^{-2}$, $\alpha(M_Z)=1/127.937$, $\sin^2\theta_W=0.23126$, $M_W=80.433 $ GeV and $M_Z=91.1876$ GeV.
We can see that both the $\lambda's$ becomes negative at the certain energy scale, the so-called instability scale $\Lambda_I$, and remains negative up to $\mpl$. One can note that for this choice of BMP, we find that the beta function of the quartic couplings $(\equiv dV(\phi)/d\phi)$ cross negative to positive around the energy scale $\sim 10^{17}$ GeV, which implies that there could be a deeper minima near that scale. We have checked that the electroweak vacuum corresponding to this point is metastable in both the Higgs and singlet scalar field directions. We also presented the corresponding effective potential along Higgs and singlet scalar field directions in Fig.~\ref{fig:s1pot} for the same benchmark points.

\begin{figure}[h!]
	\centering
	\includegraphics[scale=.34]{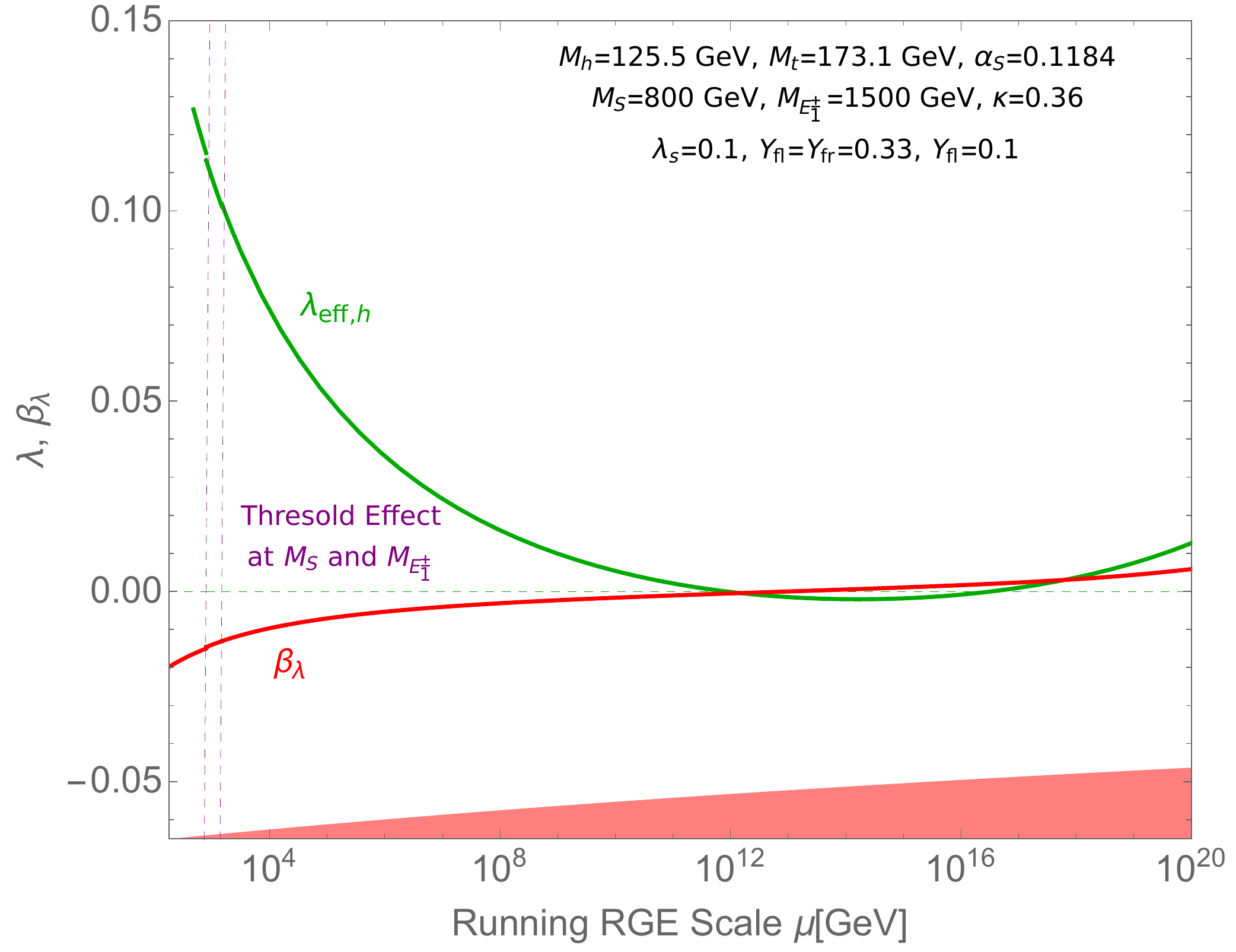}
	\includegraphics[scale=.33]{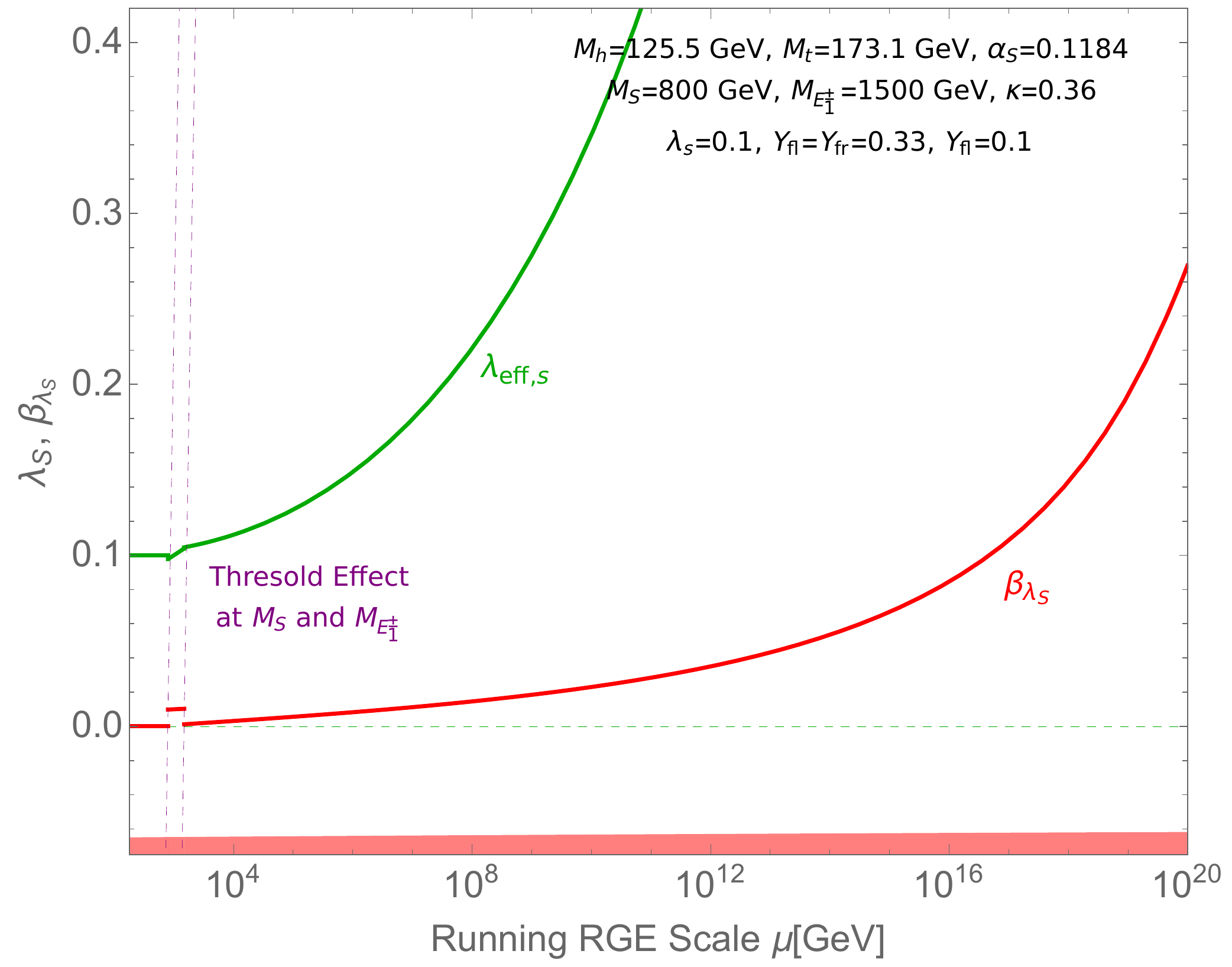}
	\caption{\it  RG evolution of the Higgs and singlet scalar quartic couplings. These plots left side for the running of $\lambda$ whereas right side for $\lambda_S$ with the choice of parameters $M_h=125.5$ GeV, $M_t=173.1$ GeV, $\alpha_S=0.1184$ and $M_{E_1^\pm}=1500$ GeV, $M_{E_2^\pm}=3000$ GeV with $\kappa=0.36$, $Y_{f}=Y_{f}^\prime$=0.33 and $Y_{N}=0.1$.}	\label{fig:s1a}
\end{figure}
\begin{figure}[h!]
	\centering
	\includegraphics[scale=.34]{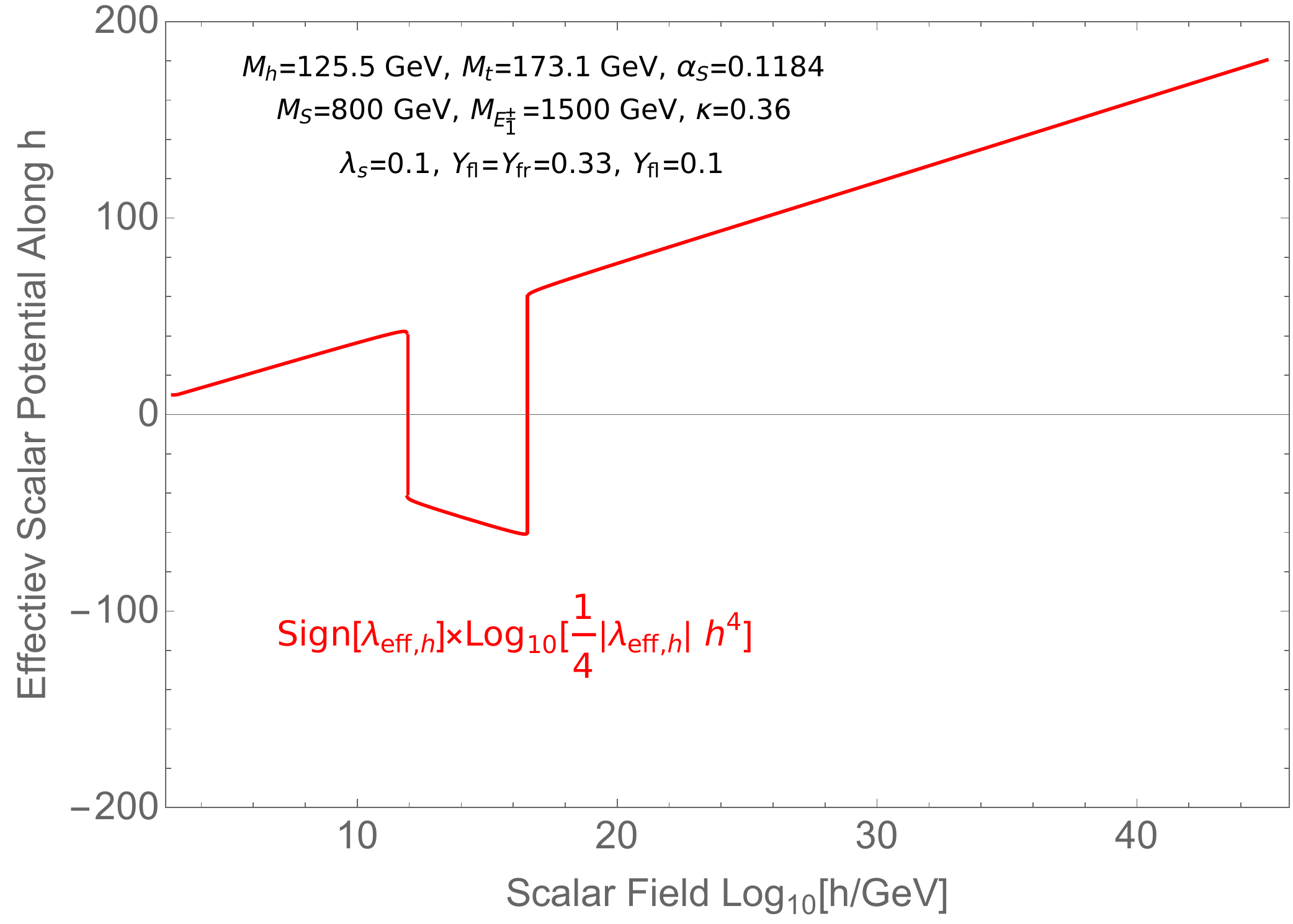}
	\includegraphics[scale=.34]{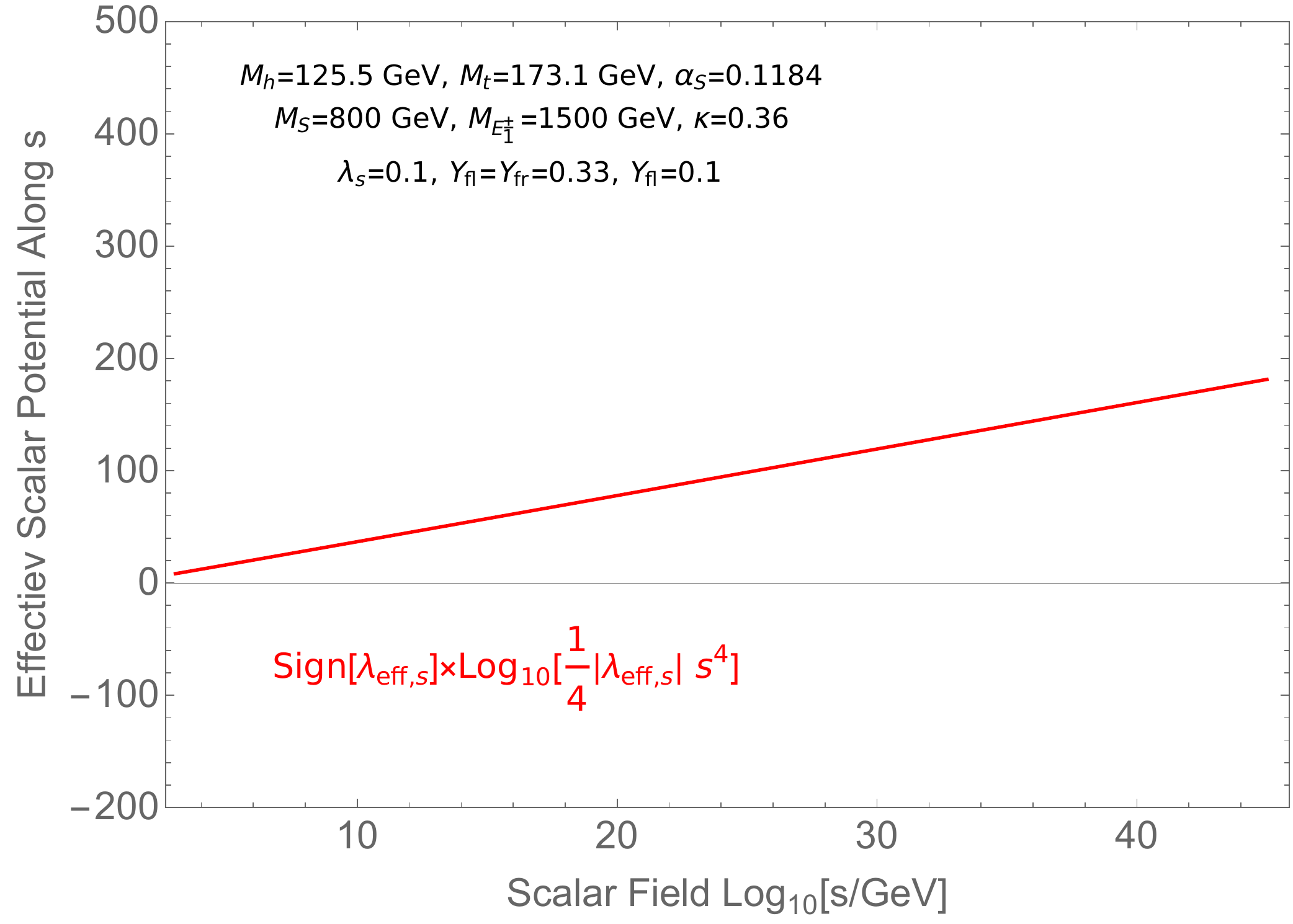}
	\caption{\it The effective potential for the choice of parameters $M_h=125.5$ GeV, $M_t=173.1$ GeV, $\alpha_S=0.1184$ and $M_{S}=800$ GeV, $M_{E_1^\pm}=1500$ GeV, $M_{E_2^\pm}=3000$ GeV with $\kappa=0.36$, $Y_{f}=Y_{f}^\prime$=0.33 and $Y_{N}=0.1$.}	\label{fig:s1pot1}
\end{figure}

Similarly, in the second Fig.\ref{fig:s1a}, we show the running of the quartic coupling for a different value of $\kappa=0.36$, keeping fixed the other parameters. Corresponding effective potential are shown in Fig.~\ref{fig:s1pot1}. There are no minima along the singlet scalar direction. However, along the Higgs direction, a minima resides near the Planck scale, and the depth of the minima is deeper than the electroweak one again; hence transition is still possible in this direction. We find the tunneling probability is much smaller than the previous choice of BMP with $\kappa=0.18$.
We want to point out that this choice of BMP with large  $\kappa=0.36$ and dark matter mass $800$ GeV violates the direct detection cross-section and over closes the Universe $\Omega h^2>1$.

\subsection{Phage diagrams}

In this model, the Higgs portal coupling $\kappa$ contributes positively to the running of Higgs quartic coupling $\lambda$. In contrast, the new Yukawa coupling $Y_N$ puts a negative contribution to the running of $\lambda$ (see Appendix~\ref{sec:app} for details). Due to additional doublet fermions, the RGEs $SU(2)$ gauge coupling $g_2$ has also been modified here, which also affects only at the threshold fermion Mass. This effect can, in turn, positively contribute to the running of $\lambda$.
The additional Yukawa couplings $Y_f$ and $Y_f^\prime$ appear at the 2-loop only. The contributions could be negative and/or positive depending on the parameters and the running effect of the other couplings. These effects are small but not insignificant in the tunneling probability calculations. 
Similarly, the singlet scalar quartic coupling $\lambda_S$ gets a positive contribution from the Higgs portal coupling $\kappa$. However, the other two new Yukawa couplings $Y_f$ and $Y_f^\prime$ provide a negative contribution to the running of $\lambda_S$ (see Appendix~\ref{sec:app}). On the other hand, the effect of $Y_N$ comes at two-loop only.

We want to point out to the readers the situation in the standard model again. The running of Higgs quartic coupling $\lambda$, top Yukawa coupling $Y_t$, and most importantly, the gauge couplings (mainly $SU(3)$ gauge coupling $g_3$) are responsible for having an additional high scale minima along the Higgs field direction. If we remove the effect of top Yukawa and/or $SU(3)$ gauge coupling, then there are no high scale minima. The standard model Higgs potential also becomes `unbounded from below' without the gauge couplings. Here, these gauge couplings and new Yukawa couplings $Y_f$ and $Y_f^\prime$ and Higgs portal coupling $\kappa$ are the evildoers for having another high scale minima along the singlet scalar field direction. In the absence of the $\kappa$ and gauge couplings for Yukawa couplings with values $\mathcal{O}(0.1)$, the potential becomes `unbounded from below' along the singlet scalar field direction too.

In order to show the explicit dependence of the electroweak stability for various parameters of this model, we present different kinds of phase diagrams.
Suppose this model is valid up to the Planck scale, which also provides the exact dark matter abundance of the Universe. In that case, these phage diagrams become important to realize where the present electroweak vacuum is residing.

\begin{figure}[h!]
	\centering
	\includegraphics[scale=.34]{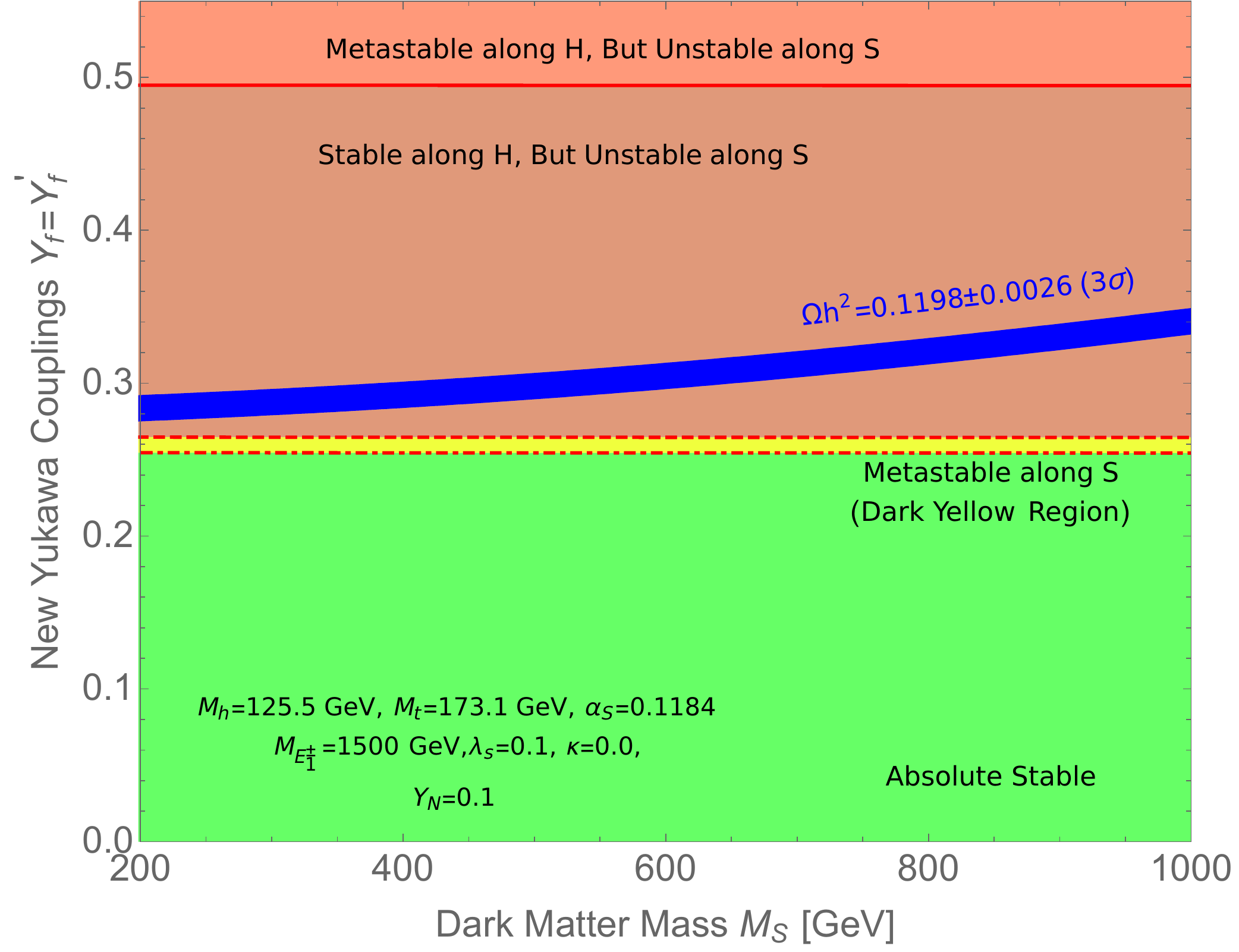}
	\includegraphics[scale=.34]{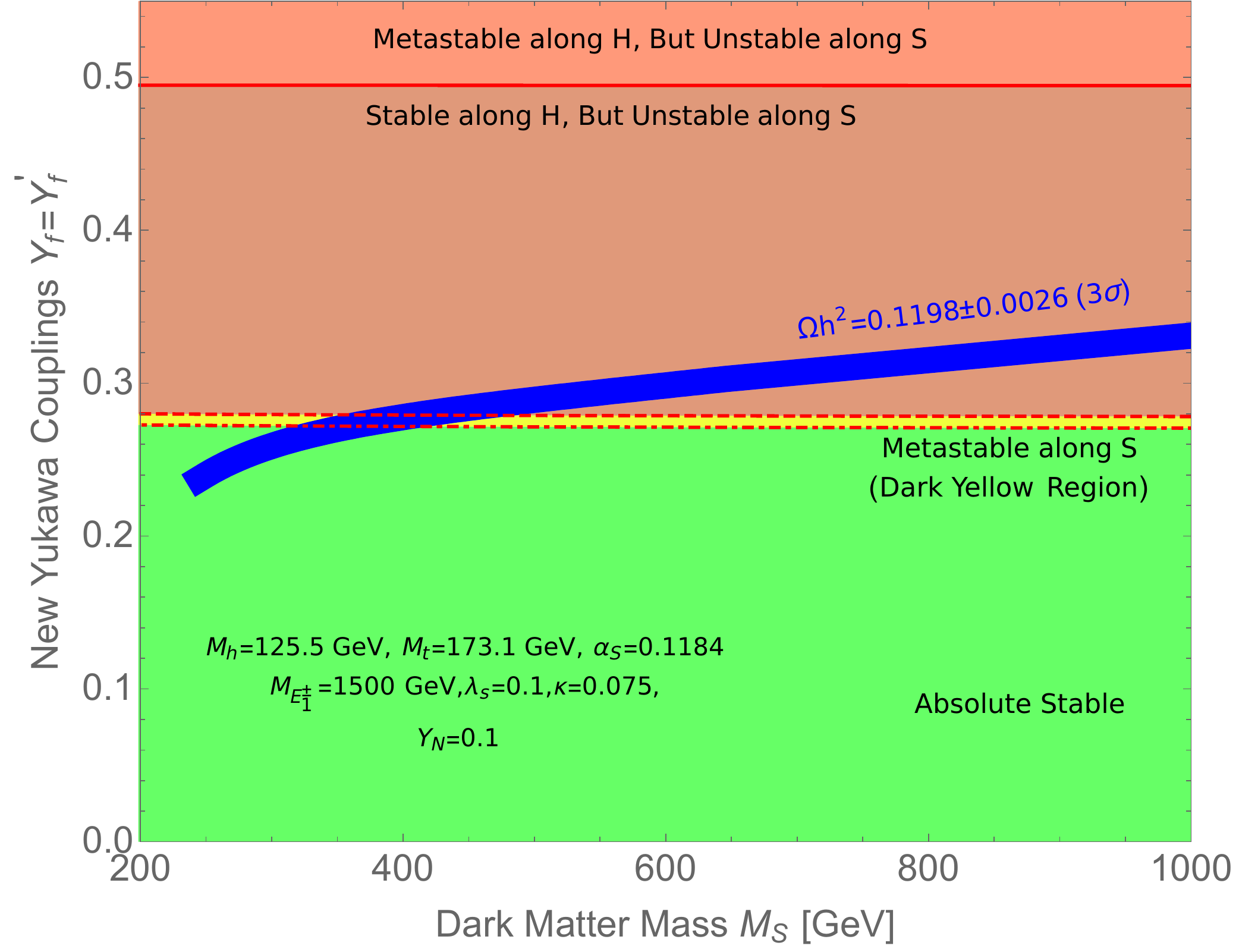}
	\caption{\it The phage diagrams in the $M_{S}-Y_N$ plane. We keep fixed the central value of the standard model parameters as discussed in the text. The green region stands for the absolute stable scalar potetial. The yelow region is the metastable region. The red region is ruled out from the instability of the scalar potential.  The left plot is drawn for $\kappa=0$ whereas the right one with $\kappa=0.075$. The blue band gives the exact relic density $\Omega h^2=0.1198\pm 0.0026$.}	\label{fig:s2}
\end{figure}
In Fig.~\ref{fig:s2}, we show the allowed parameter spaces in the $M_{S}-Y_f$ plane for central values of the standard model parameters and two different sets of new benchmark points. We use $Y_f=Y_f^\prime$, i.e., the coupling strength with the left and right-handed standard model leptons are the same. The central value of the standard model parameters are: $G_F=1.16637 \times 10^{-5}$ GeV$^{-2}$,  $\alpha(M_Z)=1/127.937$, $\alpha_S=0.1184$, $\sin^2\theta_W=0.23126$, $M_W=80.433 $ GeV, $M_Z=91.1876$ GeV, $M_h=125.5$ GeV, $M_t=173.1$ GeV. We carefully use the threshold effect at the dark matter mass $M_{S}$ and new fermion mass.
In these plots, we also keep fixed $M_{E_1^\pm}=1500$ GeV, $M_{E_2^\pm}=3000$ GeV, $Y_N=0.1$, $\lambda_S=0.1$. The left plot corresponds to Higgs portal coupling $\kappa=0$ whereas, in the right plot, we consider a different value of $\kappa=0.075$. Here the green region of these plots indicates that the scalar potential remains absolutely stable all way up to the Planck scale. The dark Yellow region (between two dashed lines) just above the green region shows the metastability region.
In this region, one can have minima along the singlet scalar directions and is deeper than the electroweak minima. We find the tunneling time $\tau_{EW}>\tau_U\approx 13.7$ billion years. It is also to be noted that there exists a minima at high scale along the Higgs field direction, but the depth of the minima is not deeper than the electroweak minima. Hence, no transition take place along the Higgs fields.
The light red shaded region above the metastability region shows the instability (remain unstable up to $\mpl$) along the singlet scalar field direction. The tunneling time here $\tau_{EW}<\tau_U$. In this region, the scalar potential along the Higgs field directions remains stable up to $Y_f=Y_f^\prime \approx 0.5$ (below the solid red lines). Above the solid red lines, i.e., $Y_f=Y_f^\prime > 0.5$, one can encounter deeper minima along the Higgs field directions. We find the tunneling time $\tau_{EW}>\tau_U$. In this region, the tunneling probability along the singlet scalar field direction is much higher than along the Higgs direction. Because the $s$ direction vacuum is deeper than the $h$ direction. Hence the universe tends to tunnel along the $s$ directing only, and here $\tau_{EW}<\tau_U$. The blue band gives the exact relic density $\Omega h^2=0.1198\pm 0.0026$.
\begin{figure}[h!]
	\centering
	\includegraphics[scale=.34]{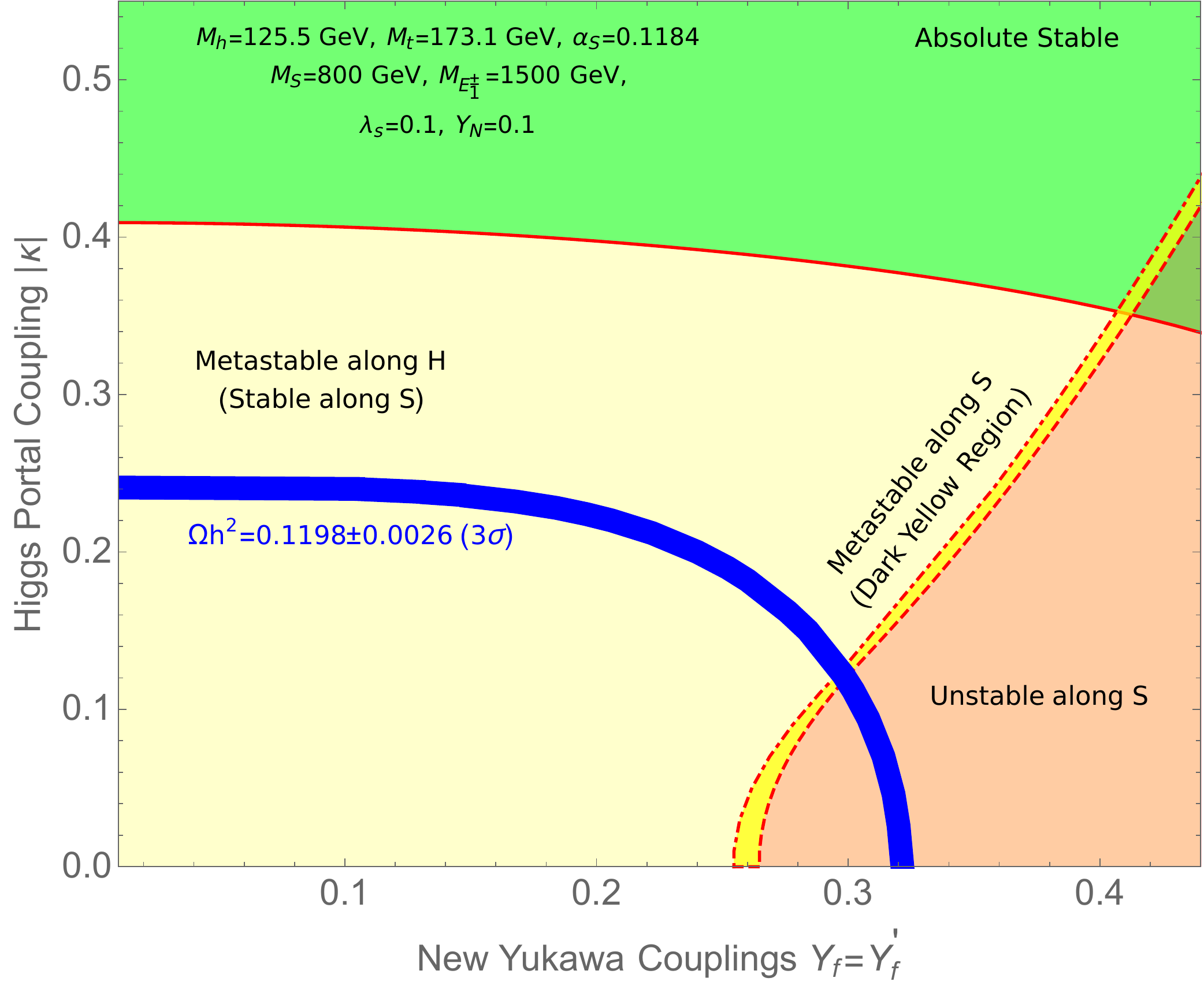}
	\includegraphics[scale=.34]{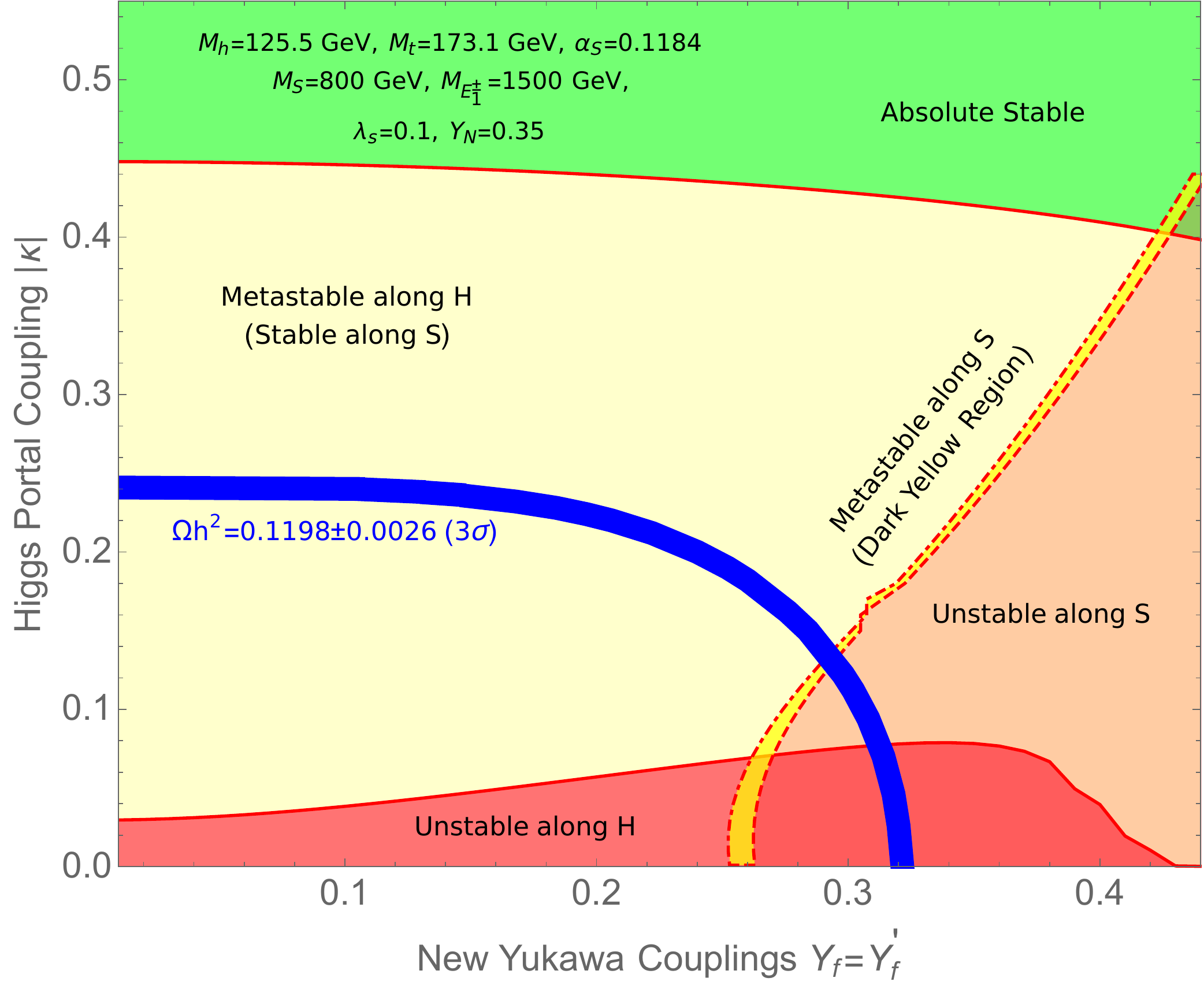}
	\caption{\it  The phage diagrams in the $\kappa-Y_f$ plane. We consider $Y_f=Y_f^\prime$ for simplicity. We keep fixed the central value of the standard model parameters as discussed in the text. The green region stands for the absolute stable scalar potetial. The yelow region is the metastable region. The red region is ruled out from the instability of the scalar potential. The blue band gives the exact relic density $\Omega h^2=0.1198\pm 0.0026$.}	\label{fig:s2a}
\end{figure}

In Fig.~\ref{fig:s2a}, we show the phage diagrams in the $\kappa-Y_f$ with $Y_f=Y_f^\prime$. The standard model parameter are taken to be fixed at the central values and we use $M_{S}=800$ GeV, $M_{E_1^\pm}=1500$ GeV, $M_{E_2^\pm}=3000$ GeV, $\lambda_S=0.1$. We chose  $Y_N=0.1$ for left plot  and  $Y_N=0.35$ in the right plot. The green region is absolutely stable all way up to the Planck scale. The dark red region in the right plot only stands for the unstable scalar potential along the Higgs field direction. In contrast, the light red region (both the plot) shows the instability (remain unstable up to $\mpl$) along the singlet scalar direction, i.e., $\tau_{EW}<\tau_U$. The dark yellow region stands for the metastability along the singlet scalar direction. The light yellow region indicates the metastability along the Higgs field direction. In the left side of the dark yellow region, the scalar potential along the singlet direction is stable but metastable along the Higgs field direction. There is only one deeper minima than electroweak minima along the Higgs field direction. The stability intersection point (top-right of the plots) of the solid and dashed (red) lines show that both the high scale minima are at the same level, and we find $\tau_{EW}=\infty$. In the intersection point of unstable lines (right plot only), we can find $\tau_{EW}=\tau_U$ for both directions. The blue band gives the exact relic density $\Omega h^2=0.1198\pm 0.0026$.
\begin{figure}[h!]
	\centering
	\includegraphics[scale=.34]{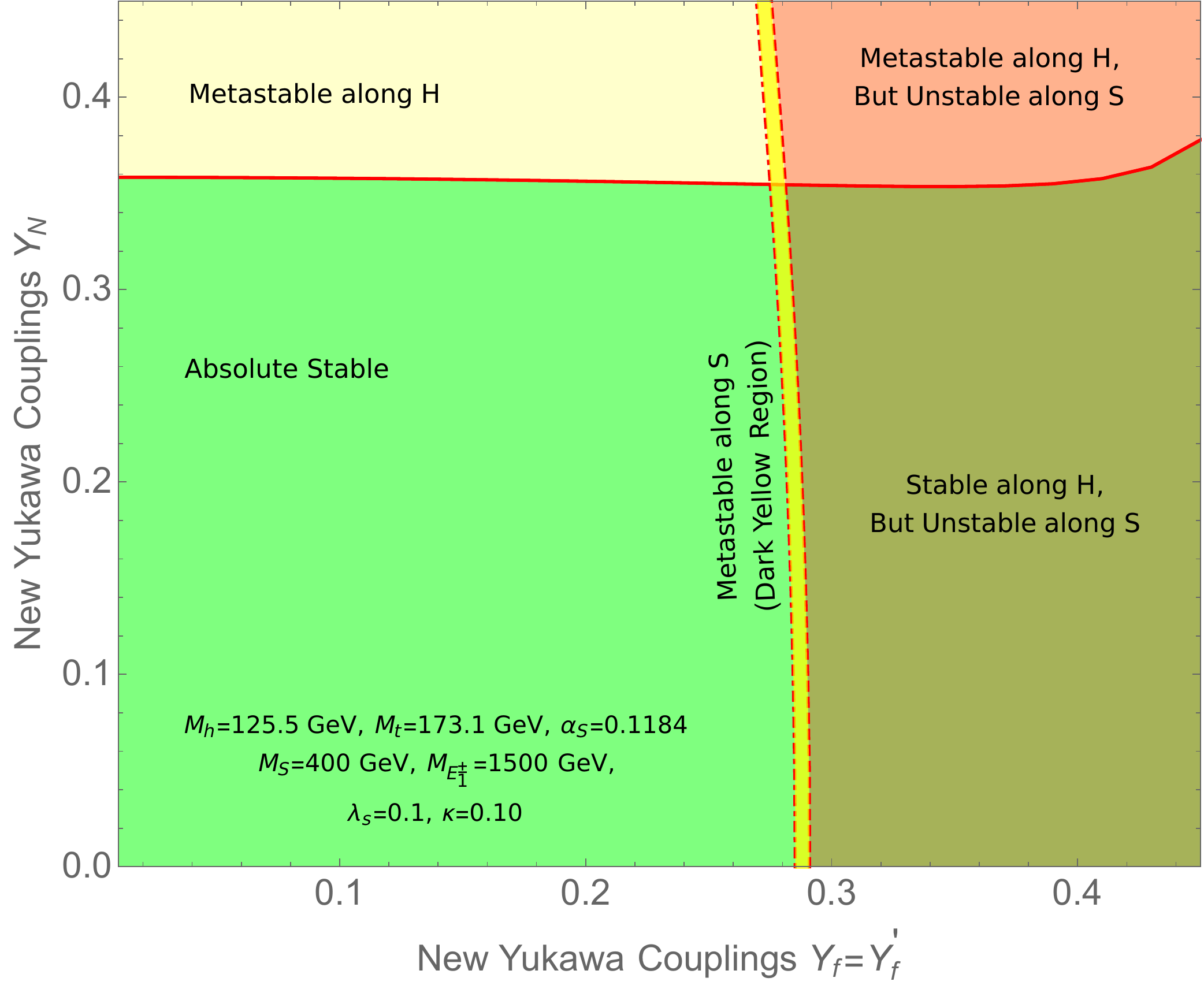}
	\caption{\it  The phage diagrams in the $Y_N-Y_f$ plane. We keep fixed the central value of the standard model parameters as discussed in the text. The green region stands for the absolute stable scalar potetial. The yelow region is the metastable region. The red region is ruled out from the instability of the scalar potential.}	\label{fig:s2b}
\end{figure}

One can see the similar effect in the $Y_N-Y_f$ plane with $Y_f=Y_f^\prime$ in Fig.~\ref{fig:s2b}. Here we fixed $M_{DM}=400$ GeV, $\kappa=0.1$ and $\lambda_S=0.1$ with $M_{E_1^\pm}=1500$ GeV, $M_{E_2^\pm}=3000$ GeV. The scalar potential is absolutely stable in the green region. In the yellowish-green region, the potential is stable along the Higgs field direction while unstable along the singlet scalar direction. The vertical dark yellow region indicates the metastability of the scalar potential along the singlet scalar field direction. The light yellow region is where the scalar potential is metastable along the Higgs field direction. Still, there is no transition along the singlet scalar direction, although the minima exists at a high scale. This minima is situated above the electroweak minima. The light red region is unstable, i.e., the tunneling time along the singlet scalar directions $\tau_{EW}<\tau_U$.
\vspace{-0.5cm}
\section{Inflation and reheating}
\label{sec:inflation}
The present data of CMB suggest the super-horizon anisotropies, measured by different experiments such as the Wilkinson Microwave Anisotropy Probe (WMAP) and Planck. It is now proved that the early universe underwent a period of rapid expansion. This rapid expansion is known as inflation. This theory of inflation can explain several cosmological problems such as the current magnetic monopole, flatness of the universe and horizon problems.
In the standard model, the Higgs quartic coupling remains negative at the GUT scale $\Lambda=10^{16}$ GeV with standard model parameters. Hence the Higgs is not a proper candidate to play the role of inflaton (see Refs.~\cite{Bezrukov:2007ep, Bezrukov:2008ut, Bezrukov:2008ej, Bezrukov:2009db} for details). Hence, we need extra new degrees of freedom to explain the inflation of the Universe~\cite{Lerner:2009xg, Lebedev:2011aq}.
As in the previous section, we have seen that one can achive $\lambda,\lambda_S>0$. Hence both the Higgs and singlets scalar (or one of them) can be considered viable inflaton candidates. We find that in the presence of varying couplings $\zeta_{h,S}\sim 1-10^5$ to Ricci scalar curvature $R$ with $\lambda,\lambda_S \approx \mathcal{O}(10^{-10}-1)$, one can easily explain the number of $e$-folds $N=30-60$ including other inflationary variables such as the tensor-to-scalar ratio $r$, the spectral index $n_s$ and the running of spectral index $n_{rs}$ which is allowed by the present experimental data~\cite{Planck:2018jri}. In this model, we use the RGEs to calculate these scalar quartic couplings at the GUT scale.

These scalars (one or two of them as inflaton) also serve as inflaton. The energy density stored in the inflaton field starts to disperse through the decay and/or annihilation into other comparatively lighter particles, including those of the SM. This epoch is known as the reheating~\cite{Allahverdi:2010xz}, which takes the Universe from the matter dominated phase during inflation to the radiation domination phase. The details discussion is shown in the Ref.~\cite{Borah:2018rca} in another context. In this model we can also achieve similar situation as in the Ref.~\cite{Borah:2018rca}.

\section{Conclusion}\label{conc}
The measurements of the Higgs properties at the Large Hadron Collider are consistent with the minimal choice of the scalar sector. However, the experimental data, including theoretical shortcomings, still allow for an extension of the standard model.
This model contains two additional vector-like charged fermions, a neutral fermion, and a real singlet scalar field. We have shown the dark matter and neutrino phenomenology in detail in the previous works~\cite{Das:2020hpd, Das:2021zea}. Both these sectors are connected through Higgs portal coupling. A brief discussion also has been added here considering the collider bound.

We have, however, seen that the singlet scalar quartic coupling $\lambda_S$ has not been entered in any physical phenomenon. The dark matter and neutrino sector, at least at tree-level, is independent of $\lambda_S$. On the other hand, the Higgs portal coupling $\kappa$ is strictly bounded by the dark matter direct detection limit(s). Depending on the dark matter mass, one can take up to $\mathcal{O}(1)$. Similarly, the new Yukawa couplings associated with vector fermions can have values $\mathcal{O}(1)$. 
Other decays/ interactions, e.g., $\Gamma(\mu \to e\gamma)$ may put stronger bounds, however, there are various ways to overcome such bounds, as discussed in Refs~\cite{Das:2020hpd,Das:2021zea}.

In this work, we discuss the range of values these couplings can take at the electroweak scale. We have seen that, at the low energy scale, the interplay of these couplings can aggravate one or more minima along the Higgs and singlet scalar field directions at high scale, threatening
the stability of the present electroweak vacuum.
We identify parameter spaces that lead to absolute stability and metastability of the electroweak vacuum.

We already know from the literature that there could be a minima only along the Higgs field directions due to the effect of the renormalisation group evolution of gauge, Yukawa, and Higgs quartic couplings. In this model, we see a similar effect to the additional scalar field direction due to the presence of new Yukawa, Higgs portal, and gauge couplings (mainly $g_1$ and $g_2$). We calculate the tunneling probability of the present electroweak minima to both the high scale minima. We show various phase diagrams identifying regions of electroweak vacuum stability and metastability from which one can easily visualize the situation of our present Universe.

One may ask why we are not taking large Higgs portal and singlet scalar quartic couplings as $\kappa,\lambda_S \leq 0.6$, which remain perturbative up to the Planck scale. It is to be noted that there could also be minima along the scalar ($h$ and $s$) fields' directions. The depth could vary depending on the parameters. 
However, the large choice of $\kappa$ will violate the direct detection bounds, as discussed in the dark matter section.
In addition, the large value of the singlet scalar quartic coupling $\lambda_S$, blows up the cannibalism effect on the dark matter in the early Universe. The dark matter density could fall rapidly due to the 4-to-2 annihilation process $SSSS \rightarrow SS$. Hence one has to avoid large singlet scalar quartic coupling in this model. The detailed discussion can be found in Ref.~\cite{Bernal:2018ins}.  
It is also to be noted that the parameters related to the FIMP dark matter do not introduce additional minima along the singlet scalar direction. The minima along the Higgs direction get modified only due to the change of gauge couplings at the threshold fermion mass.

In the summary of this work, we briefly discuss the dark matter scenario. We have done a detailed stability analysis and shown that the neutral scalar field Higgs and/or singlet scalar can play a role in the inflaton.
To the best of our knowledge, the stability analysis for this kind of model, which has two or more minima at various scale and direction, were not discussed in the literature. Hence, we present this unifying framework for simultaneously explaining dark matter, neutrino parameters, and inflation and reheating temperature at a high scale.

\section{Acknowledgements}
We would like to acknowledge support from the DAE, Government of India and the Regional Centre for Accelerator-based Particle Physics (RECAPP), HRI.

\appendix
\section{}
\label{sec:app}
In this study, we use the standard model RGEs up to three loops which
have been given in Refs.~\cite{Chetyrkin:2012rz,Zoller:2013mra,Zoller:2012cv,Chetyrkin:2013wya}. The new particle contributions are taken up to two loops which have been generated using SARAH~\cite{Staub:2013tta,Staub:2015kfa}.
In this model, the RGEs of the couplings are defined as 
\bea
\beta_{\chi_{i}}=\frac{\partial \chi_{i}}{\partial \ln \mu} =   \frac{1}{16 \pi^2}~\beta_{\chi_{i}}^{(1)}  +  \frac{1}{(16 \pi^2)^2}~\beta_{\chi_{i}}^{(2)}\, .\nn 
\eea

The RGEs of the scalar quartic couplings $\lambda_{1,2,3}$ and Yukawa couplings are given by
\vspace{-0.5cm}
\subsection{Gauge Couplings}
\vspace{-1.2cm}
{\allowdisplaybreaks  \begin{align} 
\beta_{{g'}_1}^{(1)} & =  
\frac{47}{10} {g'}_{1}^{3},~{\rm with}~ {g'}=\sqrt{5/3} g_1\\ 
\beta_{{g'}_1}^{(2)} & =  
\frac{1}{100} {g'}_{1}^{3} \Big(479 {g'}_{1}^{2} +315 g_{2}^{2} +880 g_{3}^{2} -150 Y_{N}^{2} -60 Y_{f}^2 -120 Y_{f}^{\prime \, 2} \nonumber -170 Y_t^2 \Big)\\ 
\beta_{g_2}^{(1)} & =  
-\frac{17}{6} g_{2}^{3} \\ 
\beta_{g_2}^{(2)} & =  
\frac{1}{60} g_{2}^{3} \Big( -30 Y_{N}^{2}  + 595 g_{2}^{2}  -60 Y_{f}^2  + 63 {g'}_{1}^{2}  + 720 g_{3}^{2}  -90 Y_{t}^2 \Big)\\ 
\beta_{g_3}^{(1)} & =  
-7 g_{3}^{3} \\ 
\beta_{g_3}^{(2)} & =  
-\frac{1}{10} g_{3}^{3} \Big(-11 {g'}_{1}^{2} + 20 Y_{t}^2  + 260 g_{3}^{2}  -45 g_{2}^{2} \Big)
\end{align}} 

\vspace{-1.3cm}
\subsection{Quartic scalar couplings}
\vspace{-1.3cm}
{\allowdisplaybreaks  \begin{align} 
\beta_{ \lambda_S}^{(1)} & =  
16  \lambda_S Y_{f}^2  + 3 \Big(4  \kappa^{2}  +  \lambda_S^{2}\Big) -48 Y_{f}^{\prime \, 4}  + 8  \lambda_S Y_{f}^{\prime \, 2} -96 Y_{f}^4\\ 
\beta_{ \lambda_S}^{(2)} & =  
\frac{72}{5} {g'}_{1}^{2}  \kappa^{2} -\frac{17}{3}  \lambda_S^{3} +72 g_{2}^{2}  \kappa^{2} -20  \lambda_S  \kappa^{2} -48  \kappa^{3} -24  \kappa^{2} Y_{N}^{2} \nonumber \\
&+\Big(56  \lambda_S  -288 g_{2}^{2}+ 96 Y_{N}^{2}  -\frac{288}{5} {g'}_{1}^{2} \Big)Y_{f}^4  +768 Y_{f}^6  +24 {g'}_{1}^{2}  \lambda_S Y_{f}^{\prime \, 2} \nonumber \\
&-12  \lambda_S^{2} Y_{f}^{\prime \, 2} -12  \lambda_S Y_{N}^{2} Y_{f}^{\prime \, 2} -\frac{576}{5} {g'}_{1}^{2} Y_{f}^{\prime \, 4} +28  \lambda_S Y_{f}^{\prime \, 4} \nonumber \\
&+12 Y_{f}^2 \Big(16 Y_{N}^{2} Y_{f}^{\prime \, 2}   +  \lambda_S ( 5 g_{2}^{2}  -2  \lambda_S - Y_{N}^{2}  + {g'}_{1}^{2})\Big)\nonumber \\
&+96 Y_{N}^{2} Y_{f}^{\prime \, 4} +384 Y_{f}^{\prime \, 6} -72  \kappa^{2}Y_{t}^2   \\ 
\beta_{ \kappa}^{(1)} & =  
12 \lambda  \kappa + \lambda_S  \kappa +4  \kappa^{2} +2  \kappa Y_{N}^{2} -\frac{9}{10} {g'}_{1}^{2}  \kappa -\frac{9}{2} g_{2}^{2}  \kappa \nonumber \\
&+8 (\kappa- Y_{N}^{2})Y_{f}^{2} +4  \kappa Y_{f}^{\prime \, 2} -8 Y_{N}^{2} Y_{f}^{\prime \, 2} +6  \kappa Y_{t}^2 \\ 
\beta_{ \kappa}^{(2)} & =  
\frac{1851}{400} {g'}_{1}^{4}  \kappa +\frac{9}{8} {g'}_{1}^{2} g_{2}^{2}  \kappa -\frac{125}{16} g_{2}^{4}  \kappa +\frac{72}{5} {g'}_{1}^{2} \lambda  \kappa +72 g_{2}^{2} \lambda  \kappa \nonumber \\
&-60 \lambda^{2}  \kappa -\frac{5}{6}  \lambda_S^{2}  \kappa +\frac{3}{5} {g'}_{1}^{2}  \kappa^{2} +3 g_{2}^{2}  \kappa^{2} -72 \lambda  \kappa^{2} -6  \lambda_S  \kappa^{2} -\frac{21}{2}  \kappa^{3} \nonumber \\ 
 &+\frac{15}{4} {g'}_{1}^{2}  \kappa Y_{N}^{2} +\frac{15}{4} g_{2}^{2}  \kappa Y_{N}^{2} -24 \lambda  \kappa Y_{N}^{2} -4  \kappa^{2} Y_{N}^{2} -\frac{9}{2}  \kappa Y_{N}^{4}  \nonumber \\ 
 &-4 (\kappa-9 Y_{N}^{2} ) Y_{f}^{4}+\frac{144}{25} {g'}_{1}^{4} Y_{f}^{\prime \, 2} +12 {g'}_{1}^{2}  \kappa Y_{f}^{\prime \, 2} -4  \lambda_S  \kappa Y_{f}^{\prime \, 2} -8  \kappa^{2} Y_{f}^{\prime \, 4}  \nonumber \\ 
 &-\frac{72}{5} {g'}_{1}^{2} Y_{N}^{2} Y_{f}^{\prime \, 2}+\frac{17}{2}  \kappa Y_{N}^{2} Y_{f}^{\prime \, 2} +28 Y_{N}^{4} Y_{f}^{\prime \, 2} -2  \kappa Y_{f}^{\prime \, 4} +36 Y_{N}^{2} Y_{f}^{\prime \, 4} \nonumber \\ 
 &+Y_{f}^{2} \Big(\frac{72}{25} {g'}_{1}^{4} +4 g_{2}^{4} +6 {g'}_{1}^{2}  \kappa +30 g_{2}^{2}  \kappa -8  \lambda_S  \kappa -16  \kappa^{2} -\frac{36}{5} {g'}_{1}^{2} Y_{N}^{2}  \nonumber \\ 
 &-12 g_{2}^{2} Y_{N}^{2} +\frac{17}{2}  \kappa Y_{N}^{2} +32 Y_{N}^{4} +40 Y_{N}^{2} Y_{f}^{\prime \, 2} \Big)+\frac{17}{4} {g'}_{1}^{2}  \kappa Y_{t}^{2} \nonumber \\ 
 &+\frac{45}{4} g_{2}^{2}  \kappa Y_{t}^{2} +40 g_{3}^{2}  \kappa Y_{t}^{2} -72 \lambda  \kappa Y_{t}^{2} -12  \kappa^{2} Y_{t}^{2} -\frac{27}{2}  \kappa Y_{t}^{4} \\ 
\beta_{\lambda}^{(1)} & =  
+\frac{27}{200} {g'}_{1}^{4} +\frac{9}{20} {g'}_{1}^{2} g_{2}^{2} +\frac{9}{8} g_{2}^{4} -\frac{9}{5} {g'}_{1}^{2} \lambda -9 g_{2}^{2} \lambda +24 \lambda^{2} \nonumber \\ 
 &+\frac{1}{2}  \kappa^{2} +4 \lambda Y_{N}^{2} -2 Y_{N}^{4}  +12 \lambda Y_{t}^{2} -6 Y_{t}^{4} \\ 
\beta_{\lambda}^{(2)} & =  
-\frac{3843}{2000} {g'}_{1}^{6} -\frac{1821}{400} {g'}_{1}^{4} g_{2}^{2} -\frac{61}{16} {g'}_{1}^{2} g_{2}^{4} +\frac{289}{16} g_{2}^{6} +\frac{2067}{200} {g'}_{1}^{4} \lambda \nonumber \\ 
 &+\frac{117}{20} {g'}_{1}^{2} g_{2}^{2} \lambda -\frac{53}{8} g_{2}^{4} \lambda +\frac{108}{5} {g'}_{1}^{2} \lambda^{2} +108 g_{2}^{2} \lambda^{2} -312 \lambda^{3} \nonumber \\ 
 &-5 \lambda  \kappa^{2} -2  \kappa^{3} -\frac{9}{4} {g'}_{1}^{4} Y_{N}^{2} +\frac{33}{10} {g'}_{1}^{2} g_{2}^{2} Y_{N}^{2} -\frac{3}{4} g_{2}^{4} Y_{N}^{2} +\frac{15}{2} {g'}_{1}^{2} \lambda Y_{N}^{2} \nonumber \\ 
 &+\frac{15}{2} g_{2}^{2} \lambda Y_{N}^{2} -48 \lambda^{2} Y_{N}^{2} -\frac{12}{5} {g'}_{1}^{2} Y_{N}^{4} - \lambda Y_{N}^{4} +10 Y_{N}^{6} \nonumber \\ 
 &+\Big(2 Y_{N}^{4}  -3 \lambda Y_{N}^{2}  -4  \kappa^{2} \Big)Y_{f}^{2} -2  \kappa^{2} Y_{f}^{\prime \, 2} -3 \lambda Y_{N}^{2}  Y_{f}^{\prime \, 2} +2 Y_{N}^{4}  Y_{f}^{\prime \, 2} \nonumber \\ 
 & -\frac{171}{100} {g'}_{1}^{4}  Y_{t}^{2} +\frac{63}{10} {g'}_{1}^{2} g_{2}^{2}  Y_{t}^{2} -\frac{9}{4} g_{2}^{4} Y_{t}^{2}  +\frac{17}{2} {g'}_{1}^{2} \lambda Y_{t}^{2}  +\frac{45}{2} g_{2}^{2} \lambda Y_{t}^{2}  \nonumber \\ 
 &+80 g_{3}^{2} \lambda Y_{t}^{2}  -144 \lambda^{2} Y_{t}^{2} -\frac{8}{5} {g'}_{1}^{2} Y_{t}^{4} -32 g_{3}^{2} Y_{t}^{4}  -3 \lambda Y_{t}^{4}  +30 Y_{t}^{6} 
\end{align}} 
\vspace{-1.3cm}
\subsection{Yukawa Couplings}
\vspace{-1.3cm}
{\allowdisplaybreaks  \begin{align} 
\beta_{Y_t}^{(1)} & =  
\frac{3}{2}  Y_{t}^{3}+Y_t \Big( 3  Y_{t}^{2}  -8 g_{3}^{2}  -\frac{17}{20} {g'}_{1}^{2}  -\frac{9}{4} g_{2}^{2}  + Y_{N}^{2} \Big)\\ 
\beta_{Y_t}^{(2)} & =  
+\frac{1}{80} \Big(120  Y_{t}^{5}+ Y_{t}^{3} (1280 g_{3}^{2}  -180 Y_{N}^{2}  + 223 {g'}_{1}^{2}   -540  Y_{t}^{2}  \nonumber \\ 
 &+ 675 g_{2}^{2}  -960 \lambda ) \Big)+Y_t \Big(\frac{1361}{600} {g'}_{1}^{4} -\frac{9}{20} {g'}_{1}^{2} g_{2}^{2} -\frac{11}{2} g_{2}^{4} +\frac{19}{15} {g'}_{1}^{2} g_{3}^{2} \nonumber \\ 
 &+9 g_{2}^{2} g_{3}^{2} -108 g_{3}^{4} +6 \lambda^{2} +\frac{1}{4}  \kappa^{2} +\frac{15}{8} {g'}_{1}^{2} Y_{N}^{2} +\frac{15}{8} g_{2}^{2} Y_{N}^{2} -\frac{9}{4} Y_{N}^{4} \nonumber \\ 
 &-\frac{3}{4} Y_{N}^{2} Y_{f}^{2} -\frac{3}{4} Y_{N}^{2} Y_{f}^{\prime \, 2} +\frac{17}{8} {g'}_{1}^{2} Y_{t}^{2}  +\frac{45}{8} g_{2}^{2} Y_{t}^{2}  +20 g_{3}^{2} Y_{t}^{2}  -\frac{27}{4} Y_{t}^{4}  \Big)\\ 
\beta_{Y_{f}^{\prime}}^{(1)} & =  
\Big(4 Y_{f}^{2}   + 5 Y_{f}^{\prime \, 2}   -\frac{18}{5} {g'}_{1}^{2}  + Y_{N}^{2}\Big) Y_{f}^{\prime}  \\ 
\beta_{Y_{f}^{\prime}}^{(2)} & =  
\Big(\frac{669}{50} {g'}_{1}^{4} +\frac{1}{12}  \lambda_S^{2} + \kappa^{2}-\frac{177}{40} {g'}_{1}^{2} Y_{N}^{2} +\frac{51}{8} g_{2}^{2} Y_{N}^{2} -4  \kappa Y_{N}^{2} \nonumber \\ 
 &-\frac{7}{4} Y_{N}^{4} -10 Y_{f}^{4} +Y_{f}^{2} \Big(-14 Y_{f}^{\prime \, 2} + 15 g_{2}^{2}  + 3 {g'}_{1}^{2}  + \frac{3}{4} Y_{N}^{2} \Big)\nonumber \\ 
 &+\frac{102}{5} {g'}_{1}^{2} Y_{f}^{\prime \, 2} -2  \lambda_S Y_{f}^{\prime \, 2} -\frac{21}{4} Y_{N}^{2}  Y_{f}^{\prime \, 2}  -\frac{57}{4} Y_{f}^{\prime \, 4} -\frac{9}{2} Y_{N}^{2} Y_{t}^{2} \Big) Y_{f}^{\prime}\\ 
\beta_{Y_{f}}^{(1)} & =  
\frac{1}{10} \Big(20 Y_{f}^{\prime \, 2}  -45 g_{2}^{2}  + 5 Y_{N}^{2}  + 70 Y_{f}^{2}  -9 {g'}_{1}^{2} \Big)Y_{f} \\ 
\beta_{Y_{f}}^{(2)} & =  
\Big(\frac{1419}{400} {g'}_{1}^{4} -\frac{27}{40} {g'}_{1}^{2} g_{2}^{2} -\frac{521}{16} g_{2}^{4} +\frac{1}{12}  \lambda_S^{2} + \kappa^{2}\nonumber \\ 
 &-\frac{123}{80} {g'}_{1}^{2} Y_{N}^{2} +\frac{33}{16} g_{2}^{2} Y_{N}^{2} -2  \kappa Y_{N}^{2} - Y_{N}^{4} -\frac{105}{4} Y_{f}^{4} \nonumber \\ 
 &-Y_{f}^{2} \Big( 33 g_{2}^{2} -2  \lambda_S -7 Y_{f}^{\prime \, 2}  + \frac{33}{5} {g'}_{1}^{2}  -\frac{33}{8} Y_{N}^{2} \Big)+6 {g'}_{1}^{2} Y_{f}^{\prime \, 2} \nonumber \\ 
 &-\frac{9}{8} Y_{N}^{2} Y_{f}^{\prime \, 2} -5 Y_{f}^{\prime \, 4}  -\frac{9}{4} Y_{N}^{2} Y_{t}^{2} \Big)Y_{f}\\ 
\beta_{Y_{N}}^{(1)} & =  
\frac{1}{4} \Big(2 Y_{f}^{2} + 10 Y_{N}^{2} + 12 Y_{t}^{2}  + 2 Y_{f}^{\prime \, 2}  -9 {g'}_{1}^{2}  -9 g_{2}^{2} \Big)Y_{N}\\ 
\beta_{Y_{N}}^{(2)} & =  
\frac{1569}{200} {g'}_{1}^{4} Y_{N} +\frac{27}{20} {g'}_{1}^{2} g_{2}^{2} Y_{N} -\frac{11}{2} g_{2}^{4} Y_{N} +6 \lambda^{2} Y_{N}  \nonumber \\ 
 &+\frac{1}{4}  \kappa^{2} Y_{N} +\frac{537}{80} {g'}_{1}^{2} Y_{N}^{3} +\frac{165}{16} g_{2}^{2} Y_{N}^{3} -12 \lambda Y_{N}^{3} -3 Y_{N}^{5} \nonumber \\ 
 &-\frac{25}{8} Y_{N}  Y_{f}^{4} -3 {g'}_{1}^{2} Y_{N}  Y_{f}^{\prime \, 2} -2  \kappa Y_{N}  Y_{f}^{\prime \, 2} -\frac{7}{8} Y_{N}^{3}  Y_{f}^{\prime \, 2} \nonumber \\ 
 &-\frac{13}{8} Y_{N}  Y_{f}^{\prime \, 4} -\frac{1}{10}  Y_{f}^{2} Y_{N} \Big(10 Y_{N}^{2}  + 15 g_{2}^{2}  + 20  \kappa  + 21 {g'}_{1}^{2}  + 45 Y_{f}^{\prime \, 2} \Big) \nonumber \\ 
 & +\frac{17}{8} {g'}_{1}^{2} Y_{N} Y_{t}^{2} +\frac{45}{8} g_{2}^{2} Y_{N} Y_{t}^{2}  +20 g_{3}^{2} Y_{N} Y_{t}^{2}  -\frac{27}{4} Y_{N}^{3} Y_{t}^{2}  -\frac{27}{4} Y_{N} Y_{t}^{4}  
\end{align}} 
\vspace{-1.3cm}
\subsection{$\gamma$ functions}
\vspace{-1.3cm}
{\allowdisplaybreaks  
\bea
\gamma_s &=&\frac{1}{16 \pi^2} \left( 4 \, Y_f^2 + 2\, Y_f^{\prime, 2} \right) +\frac{1}{256 \pi^4} \Big(     \kappa^2 + \frac{\lambda_S^2}{12} - 10 Y_f^{\prime, 4} \nonumber\\
&&+ 10 g_1^2 Y_f^{\prime, 2} - 5 Y_f^{\prime, 4} -  3 Y_f^{\prime, 2} Y_N^2 + 3 Y_f^{\prime, 2} (5/3 g_1^2 + 5 g_2^2 - Y_N^2)    \Big) \\
\gamma_{h}^{extra} &=& \frac{ Y_N^2}{16 \pi^2} + \frac{1}{265 \pi^4} \Big(  \frac{\kappa^2}{4} + \frac{25}{8} g_1^2 Y_N^2 + \frac{15}{8} g_2^2 Y_N^2 - \frac{3}{4} Y_f^2 Y_N^2  \nonumber\\
&&- \frac{3}{4} Y_f^{\prime, 2} Y_N^2 - \frac{9}{4} Y_N^4 \Big) 
\eea
}


\newpage

\bibliographystyle{apsrev4-1}
\bibliographystyle{utphys}
\bibliography{tevportalnew}
\end{document}